# Multichannel Phonon Interconversion Drives 90° Polarization Rotation for Stable Domain Walls in Hafnium Oxide

Fatoye Sawyerr, [1,2,3†] Hamoon Fahrvandi, [4,5†] Pawan Kumar, [4,5] Zekun Zhang,[1,2,3] Shuning Lv,[2,6]
Yue Yu,[1,2,3] Xuan Gao,[1,2,3] Kang Jia,[1,2,3] Shifeng Wen,[2] Zheng Wen,[7] Li-Min Liu,[6]
Xiaoli Fan,[1,2,3]* Jun Hee Lee,[4,5]* and Tengfei Cao [1,2,3]*

*[1]Research Center for Advanced Lubrication and Sealing Materials, School of Materials Science and Engineering, Northwestern Polytechnical University, Xi'an, Shaanxi 710072, China.*

*[2]Department of Materials Science and Engineering, Northwestern Polytechnical University, Xi'an 710072, China.*

*[3]China-Kazakhstan Belt and Road Joint Laboratory on Materials Genome Engineering and Intelligent Science, Northwestern Polytechnical University, Xi'an 710072, China.*

*[4]Department of Energy Engineering, School of Energy and Chemical Engineering, Ulsan National Institute of Science and Technology (UNIST), Ulsan 44919, Republic of Korea.*

*[5]Department of Physics, Ulsan National Institute of Science and Technology (UNIST), Ulsan, 44919, Republic of Korea.*

*[6]School of Physics, Beihang University, Beijing 100191, China.*

*[7]College of Electronics and Information & Shandong Key Laboratory of Micro-nano Packaging and System Integration, Qingdao University, Qingdao 266071, China.*

## Abstract

Polarization rotation is a fundamental mechanism in ferroelectric materials, yet how such rotation arises microscopically in fluorite ferroelectrics and gives rise to distinct domain walls (DW) remains unresolved. Phonon modes offer a microscopic route to such structural transformations, as geometric orientations can modify their frequencies, dispersion relations, and vibrational eigenstates when their characteristic wavelengths become comparable to relevant physical dimensions. Here, by combining condensed phonon mode decomposition with first-principles calculations, we investigate the origin of 90° polarization rotation in the ferroelectric phase of $HfO_2$. Using the cubic parent phase as reference, we show that polarization rotation originates from multiple symmetry equivalent phonon modes of distinct eigenstates at different wavevectors in the Brillouin zone, whose coupling and cooperative reorientation continuously evolve to generate distinct 90° *Pca2₁* variants and switching pathways. Among the 90° variants, the lowest-energy switching pathway through the *Pmn2₁* exhibits an energy barrier comparable to, but slightly lower than, that of the 180° polarization reversal pathway through the *P4₂/nmc* phase, indicating that 90° polarization rotation is an energetically accessible alternative switching route. We further analyze 90° DW and find that frustrated phonon modes emerge at the wall center, where the coexistence and competition of condensed modes promote local reconfiguration of the lattice distortions, contributing to the elevated DW energies. Importantly, experimental observations with STEM imaging provide direct evidence of atomic mode distortions in stable 90° DW, consistent with the theoretically identified low-energy DW. These findings provide the fundamental structural mechanism underlying 90° switching dynamics, competing domain-wall physics, and macroscopic ferroelectric behavior.

## I. INTRODUCTION

Polarization rotation, in which the spontaneous polarization reorients between crystallographically equivalent directions, is a fundamental mechanism governing switching in ferroelectric materials [1]. In conventional perovskite ferroelectrics such as $BaTiO_3$ and $PbTiO_3$, symmetry breaking from a high symmetry parent phase produces multiple energetically equivalent polarization variants, enabling both 180° polarization reversal and non-180° switching, including 90° polarization rotation [2,3]. Correspondingly, 90° domain walls provide the structural boundaries between differently oriented variants and can play an active role in polarization switching through their nucleation and motion [4,5]. First-principles studies of $BaTiO_3$ have further shown that polarization switching can proceed through successive 90° reorientation processes, demonstrating that polarization rotation can provide an energetically competitive alternative to direct polarization reversal [6]. In many conventional perovskite ferroelectrics, polarization originates from a B site cation off-centering displacement relative to the surrounding oxygen octahedron. This makes polarization rotation a change in the dominant cation displacement vector. However, in other ferroelectric materials, such as the fluorite $HfO_2$, the atomic scale origin of polarization is fundamentally different; the polar $Pca2_1$ structure is characterized primarily by collective oxygen-anion displacements relative to the Hf sublattice, making polarization rotation a complex mechanism [7,8].

Since ferroelectricity discovery in hafnium oxide ($HfO_2$), it has emerged as a technologically important material owing to its robustness at the nanoscale and compatibility with complementary-metal-oxide-semiconductor (CMOS) technology [9,10]. To understand the switching dynamics, the macroscopic origin of the $Pca2_1$ phase from phonon analysis is important [11,12]. Using the unit cell (12 atoms) of cubic phase as reference, there are eight phonon modes that contribute to the formation of the $Pca2_1$ phase, these are $\Gamma_{15}^{z}$ , $X_2^{\prime x}$ , $X_5^{y}$ , $Y_5^{z}$ , $Z_5^{x}$ , $Z_5^{\prime y}$ , $Z_5^{\prime x}$, and $Y_3^{\prime y}$ correspondingly [13-16]. These oxygen and hafnia condense modes represent the collective displacement of respective atoms in the out of plane unit cell of the $Pca2_1$ phase. Amongst these modes the dynamically unstable and principal instability at $\omega = -i228$ cm$^{-1}$ of the Brillouin zone boundary $X$ point condensing at $X_2^{\prime}$ mode, produces the centrosymmetric tetragonal ($P4_2/nmc$) phase, which is the conventional intermediate structure upon cooling from the cubic phase [17,18]. Importantly, the instability is dominated largely by oxygen displacements, reflecting the unusual lattice dynamical origin of hafnia compared with conventional perovskite ferroelectrics [19]. The orthorhombic $Pca2_1$ phase stabilization involves cooperative coupling between several symmetry allowed phonon modes. In particular, several discussions on trilinear coupling in $HfO_2$ have acknowledged $\Gamma_{15}^{z}$ , $X_2^{\prime x}$ , $X_5^{y}$ to easily stabilized the $Pca2_1$ phase by lowering the energy cost through non polar coupling [20,21]. These polarization discussions have however been centered on switching proceeding predominantly through conventional 180° polarization reversal, mediated by domain wall motion, or electric field induced phase transitions.

For non-180° switching, experimental studies have established the orthorhombic $Pca2_1$ phase and revealed multiple polarization orientation variants, providing direct evidence that polarization rotation are intrinsic to the switching behavior of fluorite $HfO_2$ [22-24]. Also, theoretical studies have suggested that charged oxygen vacancies promote a fatigue resistant 90° polarization rotation pathway under [111] fields, highlighting charged defects as a key control parameter that can favor rotation over conventional 180° reversal [25]. At the mesoscale, 90° domain walls play a dual role in switching, facilitating 180° domain nucleation while impeding domain wall propagation because of their higher migration barrier, thereby influencing the overall switching pathway [26,27]. Despite this growing body of experimental and theoretical evidence, the microscopic lattice dynamical origin of 90° polarization rotation in hafnia remains unresolved. In particular, it remains unclear how the collective atomic distortions of one $Pca2_1$ variant reorganize into those of its symmetry equivalent 90° rotated counterpart, how this reorganization connects the diverse switching pathways, and how the same structural transformation determines the formation of distinct 90° domain wall configurations.

In this work, we demonstrate that 90° polarization rotation in $HfO_2$ arises from multiple symmetry equivalent phonon modes of distinct eigenvectors leading to rotated variants. Using condensed phonon mode analysis and first-principles calculations, we identify pristine 90° rotated $HfO_2$ symmetry variants of the $Pca2_1$ phase, and investigate their phonon modes as well as their microscopically different switching pathways and show that these pathways are microscopically different. We further reveal that the multichannel phonons have distinct switching pathways and amongst the variants can provide an alternative low energy pathway through a distorted polar $Pmn2_1$ intermediate structure, reducing the energetic cost of 90° switching. The lowered energy in these rotational variants can have a slight comparable energy barrier to conventional 180° polarization reversal. Extending the phonon-mediated mechanism to 90° domain walls, we constructed and analyzed eight domain walls from the 90° $Pca2_1$ variants. Although electrostatic interactions contribute to the energetics of some walls, they do not fully account for the elevated domain wall energies. Instead, the coexistence and competition of multiple condensed phonon modes at the wall centers induce phonon-mode frustration, hindering local lattice relaxation and providing an additional microscopic contribution to the energetic cost of domain wall formation. Finally, we confirm atomic scale displacements with experimental evidence with atomic-resolution scanning transmission electron microscopy (STEM) of the formation of stable 90° domains, consistent with the theoretically identified low energy domains. These results provide the microscopic origin of 90° polarization switching and domain wall formation in fluorite $HfO_2$.

## II. RESULTS

### A. Symmetry equivalent Phonon Modes of distinct Eigenstates

The ferroelectric *Pca2₁* phase and its rotated variants originate from symmetry breaking lattice distortions of the parent fluorite $Fm\bar{3}m$ structure of $HfO_2$. The reciprocal-space geometry of this cubic parent phase provides a natural framework for identifying the symmetry related instabilities underlying these distortions (see Fig. S1 [29]) [28]. In the face centered cubic (FCC) Brillouin zone, the high symmetry $X$ point lies on a square face and hosts an unstable zone boundary branch. Consequently, the cubic lattice symmetry distortion generates two equivalents $X$ point realizations, conventionally denoted $Y$ and $Z$, with wave vectors $X=\left(0,\frac{1}{2},\frac{1}{2}\right)$, $Y=\left(\frac{1}{2},0,\frac{1}{2}\right)$, $Z=\left(\frac{1}{2},\frac{1}{2},0\right)$. Although these wave vectors correspond to different orientations in reciprocal space, they are symmetry equivalent in the cubic parent structure. This can be realized using reduced reciprocal coordinates for FCC square faces described by $|k_x^*|=1$, $|k_y^*|=1$, and $|k_z^*|=1$, indicating that the sum of the wave vector should be 1. Therefore, the associated soft modes, $X_2^{\prime x}$, $Y_2^{\prime y}$, and $Z_2^{\prime z}$, constitute three equivalent distortion channels, distinguished not by their energetic stability but by the vector orientation and character of their atomic displacement patterns. These dominant zone boundary instabilities couple with other multiple symmetry equivalent modes across the Brillouin zone providing the symmetry allowed ingredients for constructing the polar *Pca2₁* distortions.

This symmetry equivalence is conveniently described in Fig. 1(a), where a single atom schematically represents the collective mode displacement of $Q_{m1}$, $Q_{m2}$, and $Q_{m3}$ where $m1$, $m2$ and $m3$ represents separate phonon modes and describe distinct collective atomic distortions modes associated with different wave vectors and phonon eigenvectors. In a cubic unit cell structure, the collective dominant oxygen distortion can be related using a set of mode amplitudes, for instance $\left\{Q_{X_2^{\prime x}}, Q_{Y_2^{\prime y}}, Q_{Z_2^{\prime z}}\right\}$, where each $Q$ denotes the amplitude of a distinct phonon mode. The symmetry equivalent states are therefore determined by the dominant condensed channel, with the corresponding mode acquiring a substantial amplitude while the other channels are suppressed.

Notably, this interconversion of phonon modes mechanism is not restricted to the soft modes; polar $\Gamma_{15}^x$, $\Gamma_{15}^y$, $\Gamma_{15}^z$ modes, zone boundary $X_5^y, X_5^z, Y_5^x, Y_5^z$, $Z_5^x, Z_5^y, Y_5^{\prime x}, Y_5^{\prime z}, Y_5^{\prime x}, X_5^{\prime y}, X_5^{\prime z}$, oxygen atom modes and $X_3^{\prime x}, Y_3^{\prime y}, Z_3^{\prime z}$ hafnia atom modes also participate in this interconversion, contributing to the formation of symmetry related rotated variants of the *Pca2₁* phase.

Importantly, these are not merely geometrically distinct displacement patterns, they are energetically degenerate because they are connected by the symmetry of the cubic parent lattice.

The corresponding energy landscapes exhibit equivalent double well potentials in Fig. 1(b), demonstrating that each mode channel independently provides an equivalent route from the centrosymmetric reference to a polar distorted state.

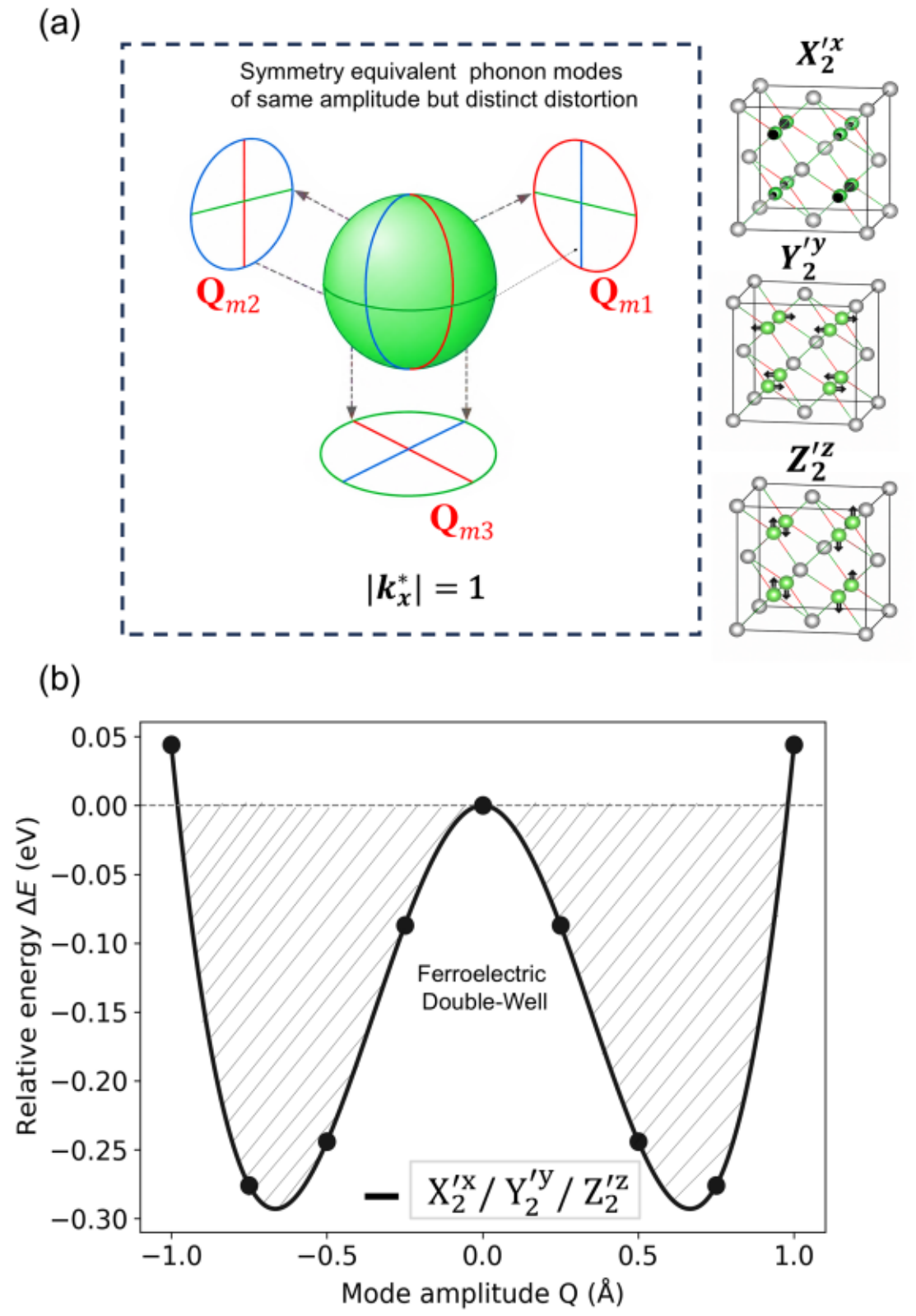

FIG. 1. Symmetry equivalent phonon modes and their distortion energy profile of cubic $HfO_2$. (a) Schematic representation of symmetry equivalent phonon modes along distinct collective direction but same mode amplitude Q, illustrating the collective distortion space spanned by the $X_2^{\prime x}$, $Y_2^{\prime y}$, and $Z_2^{\prime z}$ soft modes at the $X$ high symmetry point of the Brillouin zone. (b) Relative energy as a function of the mode amplitude Q for the symmetry equivalent $X_2^{\prime x}$, $Y_2^{\prime y}$, and $Z_2^{\prime z}$ modes.

The multichannel phonon framework provides a microscopic mechanism for polarization rotation in fluorite $HfO_2$, in contrast with conventional perovskite ferroelectrics, where symmetry equivalent eigenvectors of a single Γ-point polar instability facilitate polarization reorientation (see Fig. S2 [29]). Consequently, a change in polarization orientation is accompanied by a reorientation of the collective distortion vector in mode space, such that the dominant condensed instability can transfer between symmetry equivalent channels. Thus, the rotated *Pca2₁* variants are distinct minima associated with different directions of the multichannel distortion vector, rather than simple rigid rotations of a single real space polarization pattern. Polarization rotation thus emerges from the cooperative reorientation of symmetry equivalent phonon distortions whose eigenvectors are anchored at different but symmetry related points of the cubic Brillouin zone.

## B. Phonon modes interconversion analysis of 90° rotated *$Pca2_1$* variants of $HfO_2$

Subsequently, the origin of various polarization rotation variants can be understood by identifying the phonon modes of the ferroelectric switching process. Fig. 2 shows the atomic scale mechanism of 90° polarization rotation in ferroelectric $HfO_2$ through the evolution of symmetry equivalent orthorhombic unit cell variants. We describe the variants by $O_m^N$ where $O$ denotes the *$Pca2_1$* structure, *m* represents the polarization direction and *N* is the dominant soft mode ($X_2^{\prime x}, Y_2^{\prime y}$ and $Z_2^{\prime z}$) (see Fig. S3 [29]). Starting from the $O_z^X$ configuration, polarization rotation proceeds toward the $O_y^X$ , $O_y^Z$ , $O_x^Y$ , and $O_x^Z$ variants via a reorganization of the oxygen sublattice while the Hf framework remains nearly invariant.

The 90° rotated *$Pca2_1$* variants are connected by a collective phonon mode interconversion, where the structural distortion alternates between symmetry equivalent phonon channels without change in amplitude. The polar $\Gamma_{15}$ modes couples with the unstable soft modes and accompanying zone boundary distortions, with their eigenvector contributions being redistributed to condense specific combinations of modes that stabilize each *$Pca2_1$* variant. Each *$Pca2_1$* variant is characterized by a set of eight condensed modes which interconvert to form other rotated variant. These eight sets are $\{\Gamma_{15}^{z}, X_2^{\prime x}, X_5^{y}, Y_5^{z}, Z_5^{x}, Z_5^{\prime y}, Z_5^{\prime x}, Y_3^{\prime y}\}$, $\{\Gamma_{15}^{y}, X_2^{\prime x}, X_5^{z}, Y_5^{x}, Z_5^{y}, Y_5^{\prime x}, Y_5^{\prime z}, Z_3^{\prime z}\}$, $\{\Gamma_{15}^{y}, Z_2^{\prime z}, X_5^{y}, Y_5^{z}, Z_5^{x}, Y_5^{\prime x}, Y_5^{\prime z}, X_3^{\prime x}\}$, $\{\Gamma_{15}^{x}, Y_2^{\prime y}, X_5^{y}, Y_5^{z}, Z_5^{x}, X_5^{\prime y}, X_5^{\prime z}, Z_3^{\prime z}\}$, and $\{\Gamma_{15}^{x}, Z_2^{\prime z}, X_5^{z}, Y_5^{x}, Z_5^{y}, X_5^{\prime y}, X_5^{\prime z}, Y_3^{\prime y}\}$ corresponding to $O_z^X$, $O_y^X$, $O_y^Z$, $O_x^Y$, and $O_x^Z$ respectively.

The mode decomposition results in Fig. 2 and its detailed related amplitude in Table S1[29] reveal that each unit cell state is characterized by a distinct combination of condensed modes, demonstrating that polarization reorientation is driven by changes in alternating phonon modes but similar amplitudes. Visualization of atomic displacement patterns of the condensed modes associated with the orthorhombic *$Pca2_1$* phase of $HfO_2$ is shown in Fig. S4-S5 [29]. The plots further demonstrate that the 90° rotation pathway is heavily determined by the dominant zone boundary soft mode $X_2^{\prime x}, Y_2^{\prime y}$ and $Z_2^{\prime z}$ accompanied by alternation and redistribution of the relative contributions from the additional modes in their respective vector directions. The phonon dispersion plot of the cubic phase of $HfO_2$ path $\Gamma - Y \mid X - W - Z$ in Fig. S6[29] shows the related branches.

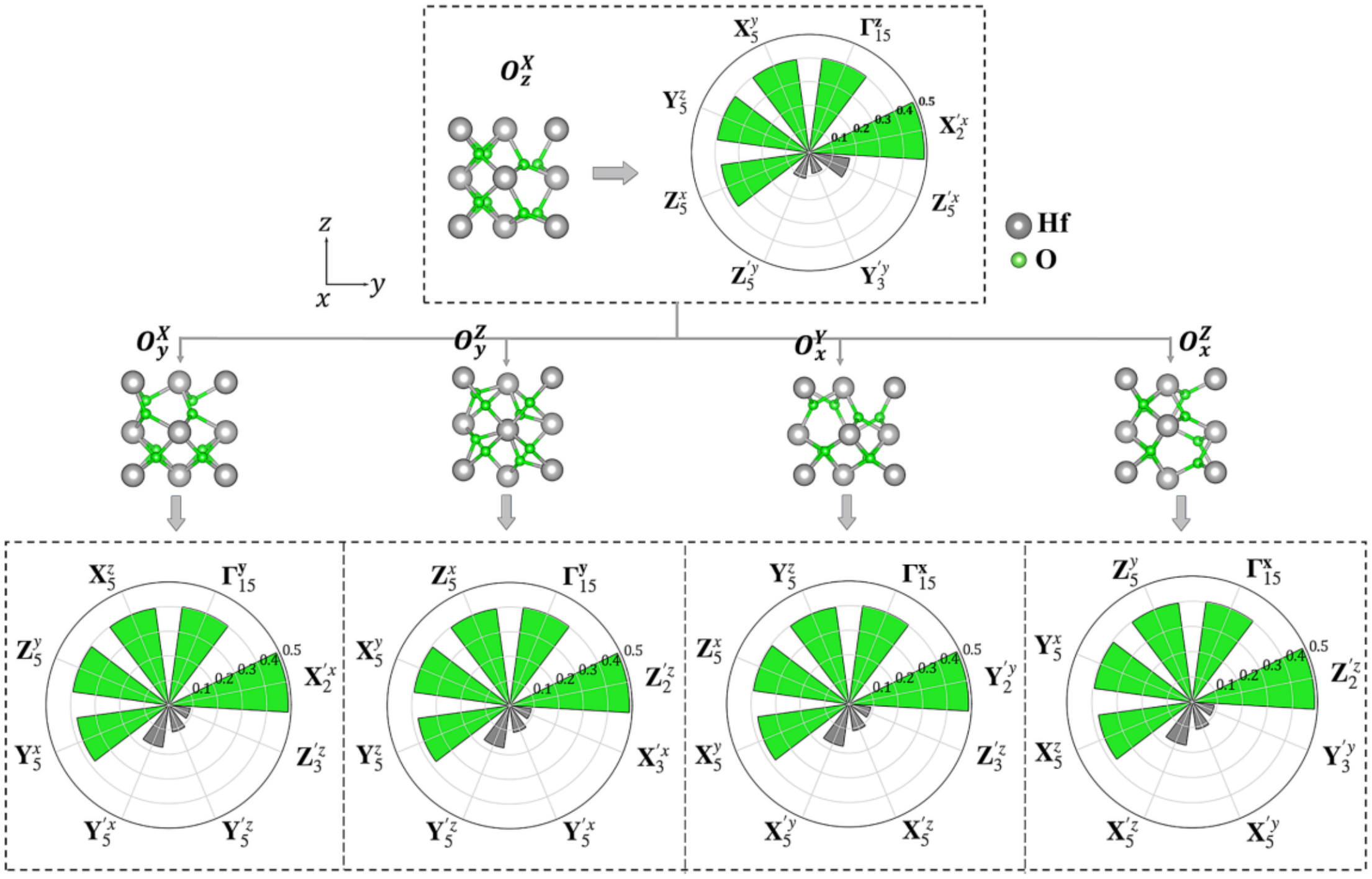

FIG 2. Phonon modes decomposition of rotated 90° ferroelectric *Pca2₁* variants of $HfO_2$. Starting from the $O_z^X$ unit cell, polarization rotates to symmetry equivalent orthorhombic variants ($O_y^X$, $O_y^Z$, $O_x^Y$, and $O_x^Z$). The accompanying mode decomposition plots show the amplitudes of the symmetry adapted lattice distortions. $O_m^N$ denotes a $Pca2_1$ structure, with polarization along the $m$-direction and soft mode of $N_2'$.

Significantly, while the overall distortion magnitude remains comparable among the symmetry related variants, the dominant mode amplitudes are selectively enhanced or suppressed depending on the final polarization direction. This behavior recognizes polarization rotation as a phonon mediated transformation between equivalent free energy minima, where the condensation and coupling of multiple lattice instabilities stabilize distinct orthorhombic variants (Fig. S7 [29]). Fig. 2 therefore provides direct evidence that 90° switching in $HfO_2$ is a collective condense mode reconfiguration driven primarily by mode interconversion.

### C. Energy landscape for phonon-mediated evolution of 90° switching in variant orthorhombic $Pca2_1$ $HfO_2$

The calculated switching energy landscapes reveal that polarization rotation in orthorhombic $HfO_2$ is intrinsically anisotropic amongst different variants. As shown in Fig. 3(a), all four 90° switching pathways exhibit individual barrier energy profiles, indicating distinct polarization rotation between equivalent ferroelectric states. Among the investigated pathways, the $O_z^X \rightarrow O_x^Y$ and $O_z^X \rightarrow O_y^Z$ transformation possesses the largest energy barrier of 0.53 eV, whereas the $O_z^X \rightarrow O_y^X$ pathway exhibits the lowest energetic cost of 0.302 eV, suggesting this path as a preferred switching mechanism. Thus, the existence of multiple energetically distinct rotational pathways demonstrates that 90° polarization switching is promoted not only by the symmetry variants but by cooperative condensed modes.

In contrast, the conventional 180° switching pathway has a comparable energy barrier of 0.31 eV passing through the non-polar tetragonal phase, $P4_2/nmc$ as shown in Fig. 3(b), highlighting a competitive switching mechanism between 90° and 180° pathway [30,31].

The energetically preferred $O_z^X \rightarrow O_y^X$ pathway proceeding through the $Pmn2_1$ intermediate phase, largely preserves the dominant condensed $X_2'^x$ oxygen distortion while requiring only a gradual redistribution of the vectors in $X_5$ mode and $\Gamma_{15}$ polar distortions. However, because the primary oxygen framework remains nearly intact, only a small lattice rearrangement is required to rotate the polarization vector, resulting in the lowest switching barrier among all four pathways. Thus, the *Pmn2₁* intermediate phase provides a structurally compatible bridge between the initial and final ferroelectric variants, allowing the polarization to rotate continuously with minimal distortion of the oxygen configuration.

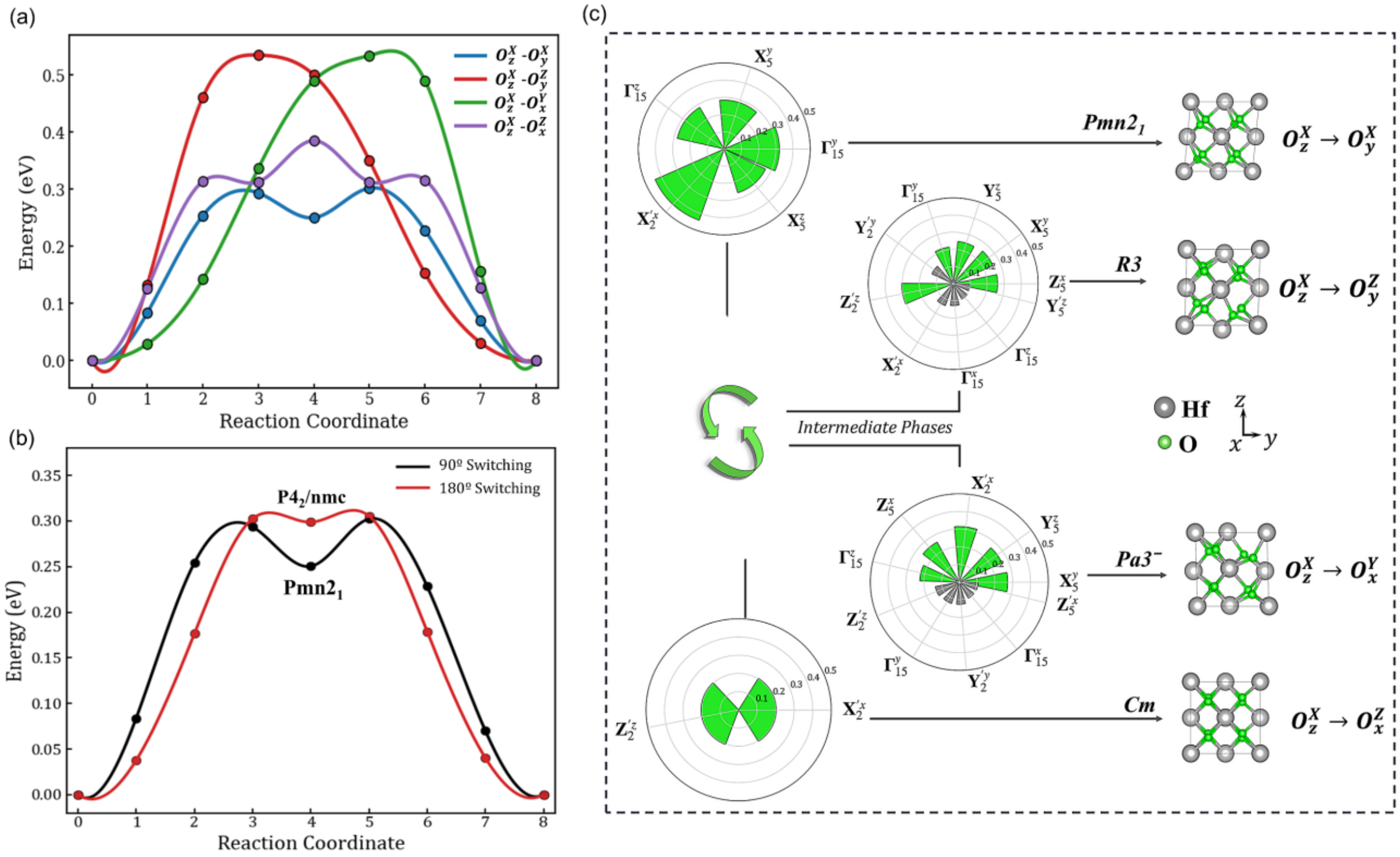

FIG 3. Energy landscapes and condense mode evolution for 90° polarization switching in orthorhombic *Pca2₁* $HfO_2$. (a) Calculated minimum energy pathways for the four symmetry-distinct 90° polarization rotation processes between orthorhombic variants ($O_z^X \rightarrow O_y^X$, $O_z^X \rightarrow O_y^Z$, $O_z^X \rightarrow O_x^Y$, and $O_z^X \rightarrow O_x^Z$,). (b) Comparison of the minimum energy pathways for 90° polarization rotation and conventional 180° polarization reversal. (c) Symmetry adapted phonon mode decomposition of the intermediate structures along the 90° switching pathways, showing the evolution of the dominant condensed modes and their corresponding crystal structures.

Also, we further analyze the distortion modes along the intermediate phases of the switching pathways in Fig. 3(c). Although all pathways originate from the same ferroelectric $O_z^X$ variant, they evolve through distinct symmetry selected intermediate structures, namely *Pmn2₁*, *R3*, *Pa*$\bar{3}$, and *Cm*, before reaching the rotated polarization state. These intermediate phases exhibit distinct profiles of condensed phonon modes, reflecting their different symmetries and underlying lattice distortions. The switching process therefore involves the selective condensation and reorganization of phonon modes as the structure evolves along each pathway. Hence, the intermediate phases represent transient geometries states along the multichannel phonon energy landscape.

## D. Phonon mode mechanism in 90° domain wall configuration and its energetics.

Unlike conventional 180° domain walls, 90° domains in $HfO_2$ possess continuous alternation and redistribution of condensed phonon modes at the wall center (see fig. S8[29]). Thus, to elucidate the microscopic origin of the distinct energetics of 90° domain walls, we constructed eight domain wall configurations connecting the different 90°*Pca2₁* variants in Fig. 4(a). Out of these eight DW, two are charged-neutral ($O_z^X/O_x^Y$, $O_z^X/O_x^Z$) and six are charged ($O_z^X/O_y^X$, $O_z^X/O_y^Z$, $O_y^X/O_x^Y$, $O_y^X/O_x^Z$, $O_y^Z/O_x^Y$, and $O_y^Z/O_x^Z$). Although electrostatic interactions contribute substantially to the excess energy of charged domain walls, an additional structural penalty arises from the differently oriented condensed phonon modes in adjacent domains. This mode mismatch induces local reorganization of the underlying lattice distortions, leading to mode competition accommodating the 90° interface.

The two equivalent variants of the 90° bidomain are moreover connected by a mixed mode motif state in which modes coexist at the wall, enabling a structurally compatible but energetically complex domain due to stress distortion at the wall. As shown in the schematic diagram in Fig. 4(b) the left domain can be characterized by the dominant mode distortion $Q_{m1}$, whereas the right domain is described by an equivalent dominant mode oriented along $Q_{m2}$or $Q_{m3}$, where Q represents the amplitude of a mode ( $m1, m2, m3$ ). At the wall center, both order parameters coexist promoting the formation of variant 90° domain walls.

This reorganization in 90° domains proceeds through a mode interconversion where the dominant mode characters of each domain progressively evolve across the wall, while modes that are suppressed become increasingly activated toward the interface. Crucially, the wall center represents a highly constrained region in which several modes coexist with substantial amplitudes. Their simultaneous activation and competition prevent the local structure from optimizing all mode contributions independently, giving rise to pronounced phonon mode frustration [32-34]. Here, we define phonon mode frustration as the incompatibility among multiple condensed phonon modes that retain substantial amplitudes at the DW center but compete to accommodate their preferred atomic displacements within the local lattice structure. This mode frustration has a direct energetic consequence, thus, if the local structural constraints force, coupled modes into a configuration that is not their preferred energetic minimum, the coupling terms and distortion energies can produce an excess energy. Studies on $HfO_2$ domain walls explicitly demonstrates that different ways of combining or reversing condensed phonon modes produce different domain wall energies, with appropriate mode coupling capable of compensating the wall energy in some configurations [35].

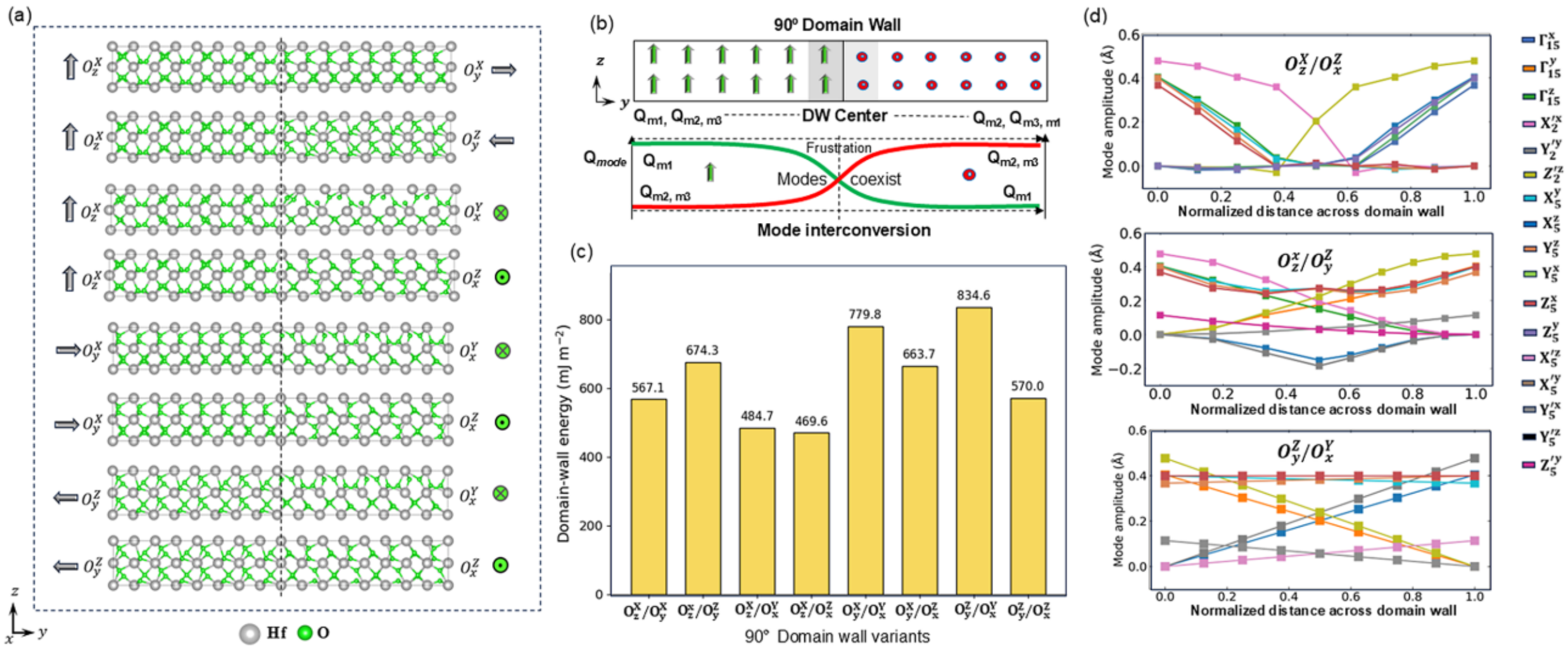

FIG 4. 90° domain wall configurations and its associated energetics in $HfO_2$. (a) Atomistic structures of 90° domain wall variants, constructed from *Pca2₁* distinct rotated variants. (b) Schematic illustration of the mode interconversion mechanism across a 90° domain wall, where multiple modes acquire substantial amplitudes and progressively interconvert. Their simultaneous activation and competition of coexisting modes produce phonon mode frustration at the wall center. Q represents the amplitude of modes $(m1, m2, m3)$ (c) Domain Wall Energies of all 90º domain structures generated from rotated variants. (d) Spatial evolution of the symmetry mode amplitudes across selected generated 90º domain walls. The units of these structural order parameters are that of distance (in Å) because they are measuring the amount of distortion.

The Domain Wall Energy (DWE) is analyzed to determines the stability and energetic cost of the 90° domain walls.

In Fig 4(c), the calculated domain wall energies (DWE) reveal a highly anisotropic energy landscape for 90° domain variants of the orthorhombic $HfO_2$. The eight 90º domain wall variants exhibit substantially different domain wall energies with, $O_z^X/O_y^X$, $O_z^X/O_y^Z$, $O_z^X/O_x^Y$, $O_z^X/O_x^Z$, $O_y^X/O_x^Y$, $O_y^X/O_x^Z$, $O_y^Z/O_x^Y$, and $O_y^Z/O_x^z$, having energies of 567.1, 674.3, 484.7, 469.6, 779.8, 663.7, 834.6 and 570.0 $mJm^{-2}$, respectively (see Table S2[29]).

The elevated DWE of charged 90° domain walls is primarily attributed to the electrostatic penalty associated with polarization discontinuities, whereas the two charge-neutral walls exhibit lesser energies of 484.7 and 469.6 $mJm^{-2}$. Nevertheless, all 90° domain walls exhibit phonon mode frustration arising from the coexistence and competition of condensed modes at the wall center, which induces local mode mismatch and contributes to the wall energy. The mode amplitude profiles across the domain wall in Fig. 4(d) and Fig. S10[29] directly reveals this distinction for all the 90° domain variants. At their domain wall center, the coexistence of zone center and zone boundary modes with substantial amplitudes indicates the influence of mode competition and local structural incompatibility. Since these modes favor different atomic displacement patterns, their simultaneous activation can prevent optimal lattice relaxation, increasing energetic cost. The resulting phonon mode frustration therefore contributes to the excess domain wall energy, with stronger mode overlap and competition generally associated with a larger energetic penalty. This trend is consistent with the $O_z^X/O_x^Z$ domain wall, which exhibits the lowest DWE and fewer coexisting modes at the wall center.

## E. Experimental formation of low energy 90° domain wall in $HfO_2$.

To establish an experimental connection with our theoretical predictions, we investigate the atomic-scale domain structure of $HfO_2$. Atomic-resolution scanning transmission electron microscopy (STEM) imaging reveals adjacent ferroelectric domains in $HfO_2$ with polarization orientations rotated by 90°, meeting at a well-defined interface in Fig. 5(a) and 5(c). The two domains correspond to *Pca2₁* variants, distinguished by the orientation of their polarization and the associated collective oxygen displacements. In the $O_z^X$ variant, the dominant $Z_2^{\prime z}$ soft mode distortion stabilizes the orthorhombic ferroelectric structure with polarization along the z-direction. In the $O_x^Z$ variant, the corresponding $X_2^{\prime x}$ distortion stabilizes the rotated structure with polarization along x-direction. Therefore, the observed interface corresponds to $O_z^X/O_x^Z$ 90° domain wall. Thus, the experimentally observed 90° domain configuration reflects the interconversion between symmetry-related distortion channels rather than a rigid rotation of an otherwise unchanged polarization state.

The atomic scale origin of this distinction is further illustrated by the local diffraction patterns and enlarged structural images in Fig. 5(b) and 5(d), respectively, together with the simulated annular-bright-field (ABF) image in Fig. 5(e).

The Fourier-transformed patterns reveal distinct sets of orthorhombic reflections in the two domains, including the ((002)o), ((201)o), and ((200)o) reflections in one orientation and the ((200)o), ((110)o), and ((020)o) reflections in the other. These differences identify the distinct crystallographic orientations of the neighboring domains and provide reciprocal-space evidence for the experimentally observed 90° reorientation. The corresponding inverse-space atomic arrangements further reveal differences in the local oxygen-sublattice configuration, connecting the crystallographic rotation to the underlying lattice distortions.

These observations provide a structural basis for interpreting the different oxygen displacement patterns associated with the two variants. Within our phonon mode framework, condensation of the $Z_2^{\prime z}$ and $X_2^{\prime x}$ modes, coupled to the corresponding polar $\Gamma_{15}^{z}$ and $\Gamma_{15}^{x}$ modes and accompanying zone-boundary distortions, generates the distinct oxygen-dominated lattice configurations. The observed domain orientations and interfacial atomic arrangement are therefore consistent with the calculated mode interconversion mechanism, in which the collective oxygen displacements reorganize cooperatively to stabilize the rotated *Pca2₁* variants.

At the interface, the two distortion patterns must be accommodated within a common lattice, requiring the local reorganization of mode amplitudes. The coexistence of modes associated with both variants introduces competing atomic displacement requirements, promoting phonon mode frustration where the local structure cannot simultaneously optimize all condensed mode contributions.

These results are consistent with our theoretical framework, in which the interconversion of distinct phonon modes enables polarization rotation and facilitates the formation of alternative domain orientations. The calculated DWE suggests that the 90° configuration is energetically accessible, though not necessarily favored. Its prevalence in experiments may also depend on strain, defects and electrostatic boundary conditions. The agreement between experimental observations and theoretical predictions thus supports the physical plausibility of the $O_z^X/O_x^Z$ 90° domain wall configuration in $HfO_2$.

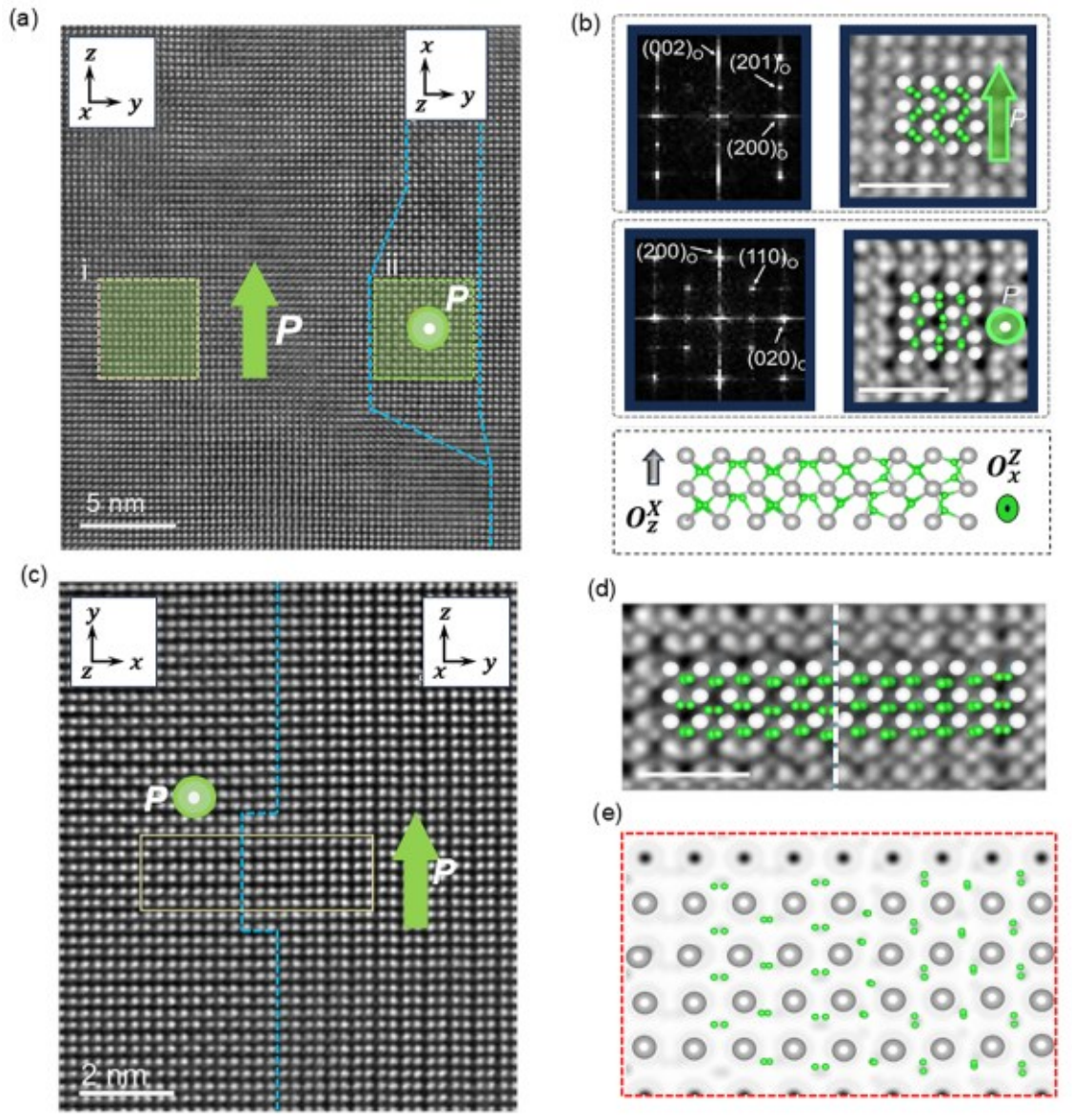

FIG 5. Experimental observation of 90° domain wall in $HfO_2$. (a) Atomic-resolution scanning transmission electron microscopy (STEM) image showing 5 nm resolution of two ferroelectric domains with regions forming 90° orientation matching to $O_z^X$ / $O_x^Z$ . (b) Corresponding Fourier-transform patterns and atomic-resolution images from the two domain regions, illustrating their distinct crystallographic orientations. Structural model illustrating $O_z^X/O_x^Z$ domain. (c) Atomic-resolution scanning transmission electron microscopy (STEM) image showing 2 nm resolution of 90° $O_z^X$ / $O_x^Z$ ferroelectric domains (d) Enlarged atomic-resolution images across the domain wall, highlighting the local atomic arrangement on either side of the interface. (e) Simulated Annular Bright Field (ABF) images of the $O_z^X/O_x^Z$ domain structure.

## III. CONCLUSION

In summary, we establish a microscopic framework for understanding 90° polarization rotation in fluorite ferroelectric $HfO_2$ by connecting condensed phonon mode evolution to polarization switching and domain wall energetics. Our results reveal that polarization rotation emerges from the cooperative reorientation and coupling of multiple symmetry equivalent phonon modes with distinct eigenstates across the Brillouin zone. Their collective evolution generates distinct *Pca2₁* variants and enables competing switching pathways through structurally and symmetrically distinct intermediate phases, including $Pmn2_1$, *R3*, *Pa3*, and *Cm*. In particular, the relatively low-energy pathway through $Pmn2_1$ demonstrates that 90° polarization rotation can compete energetically with conventional 180° polarization reversal through $P4_2/nmc$, establishing an alternative route for polarization switching in hafnia.

Extending this microscopic background to domain walls, we show that the energetic stability of 90° interfaces is determined by the interplay between local phonon mode frustration, structural compatibility, and electrostatics. The coexistence and competition of condensed modes at the wall center generate local structural distortions and contribute to the domain wall energy, whereas electrostatic interactions provide the dominant additional energetic penalty for charged configurations. These results establish a direct connection between the bulk switching landscape and the energetics of spatially inhomogeneous polarization textures. Importantly, experimental observations of stable 90° domain walls provide structural evidence consistent with our theoretically identified low-energy configurations, linking the predicted atomistic mechanisms to domain structures realized in $HfO_2$. These results establish condensed mode reorientation and phonon mode frustration as fundamental structural mechanisms driving polarization rotation and domain wall formation in fluorite ferroelectrics, providing a unified framework for understanding 90° switching, domain wall stability, and their contributions to the macroscopic behavior of $HfO_2$-based ferroelectrics.

*Acknowledgement*- This work is supported by the project of the National Natural Science Foundation of China (Program No. 12474061, 52372113,), the support from the Youth Project of Shanxi High-level Talents Introduction Plan (5113240032), the Innovation Capability Support Program of Shaanxi (Program No. 2024RS-CXTD-30). This work was also supported by the Advanced Materials-National Science and Technology Major Project (2026ZD0623501), and the funding from Suzhou National Laboratory (SK-1902-2026-004).

*These authors contributed equally: Fatoye Sawyerr & Hamoon Fahrvandi*

# Supplementary Information

# Multichannel Phonon Interconversion Drives 90º Polarization Rotation for Stable Domain Wall in Hafnium Oxide

Fatoye Sawyerr, [1,2,3†] Hamoon Fahrvandi, [4,5†] Pawan Kumar, [4,5] Zekun Zhang,[1,2,3] Shuning Lv,[2,6] Yue Yu,[1,2,3] Xuan Gao,[1,2,3] Kang Jia,[1,2,3] Shifeng Wen,[2] Zheng Wen,[7] Li-Min Liu,[6] Xiaoli Fan,[1,2,3]* Jun Hee Lee,[4,5]* and Tengfei Cao [1,2,3]*

[1]*Research Center for Advanced Lubrication and Sealing Materials, School of Materials Science and Engineering, Northwestern Polytechnical University, Xi'an, Shaanxi 710072, China.*

[2]*Department of Materials Science and Engineering, Northwestern Polytechnical University, Xi'an 710072, China.*

[3]*China-Kazakhstan Belt and Road Joint Laboratory on Materials Genome Engineering and Intelligent Science, Northwestern Polytechnical University, Xi'an 710072, China.*

[4]*Department of Energy Engineering, School of Energy and Chemical Engineering, Ulsan National Institute of Science and Technology (UNIST), Ulsan 44919, Republic of Korea.*

[5]*Department of Physics, Ulsan National Institute of Science and Technology (UNIST), Ulsan, 44919, Republic of Korea.*

[6]*School of Physics, Beihang University, Beijing 100191, China.*

[7]*College of Electronics and Information & Shandong Key Laboratory of Micro-nano Packaging and System Integration, Qingdao University, Qingdao 266071, China.*

**Computational Method**

We conducted first-principles calculations based on density-functional theory (DFT) using VASP code [1]. We adopted a generalized gradient approximation (GGA) for describing exchange correlation energy functional and pseudopotentials generated under projector-augmented planewave (PAW) scheme by Perdew-Burke-Ernzerhof for solids (PBEsol) [2], [3], [4]. Energy cut-off for plane wave basis was set to 500 eV and force tolerance for the structure optimization was 0.01 eV/Å. For $HfO_2$, (4×4×4) k-point grid was chosen under Monkhorst-Pack (MP) method for sampling integrations over the Brillouin zone [5]. The unit cell contained four formula units, orthorhombic $HfO_2$ with a = 5.2733 Å, b= 5.0551 Å and c = 5.085 Å, consistent with theoretical and experimentally measured values. The DW structure is simulated within the supercell consisting of two ferroelectric domains with different polarization orientations. Thus, the DW supercell contains 8-10 $HfO_2$ unit cells and 2 DWs. The $6 \times 1 \times 6$ and $8 \times 1 \times 8$ Monkhorst Pack k-meshes are used for the structural optimization and self-consistence calculations, respectively.

**Sample Preparation**

Sr doped $HfO_2$ thin films and ITO bottom electrodes were grown on (001)-oriented YSZ substrates by pulsed laser deposition (PLD) using a KrF excimer laser. The approximately 20-nm-thick ITO layer was deposited at a substrate temperature of 600 °C, an oxygen pressure of 0.02 mbar, a laser fluence of ~2.5 J/cm², and a pulse repetition rate of 5 Hz. Subsequently, SHO films were deposited at the same substrate temperature of 600 °C, with an oxygen pressure of 0.1 mbar, a laser fluence of ~3 J/cm², and a repetition rate of 4 Hz. Film thickness was regulated by controlling the number of laser pulses. Finally, Pt top electrodes with a diameter of approximately 50 μm and a thickness of ~30 nm were fabricated by DC magnetron sputtering through a shadow mask.

**Characterizations**

HAADF-STEM imaging was performed at an accelerating voltage of 200 kV using a spherical aberration corrected ThermoFisher Spectra 300 scanning transmission electron microscope equipped with a Super-X energy-dispersive X-ray spectroscopy (EDS) detector and a GatanContinuum 1065 spectrometer. TEM lamellae were prepared by focused ion beam milling using a Ga ion source, followed by further thinning with a precision argon ion polishing system. During HAADF-STEM acquisition, the detector collection angle was set to approximately ~50-200 mrad, with a probe convergence angle of ~21 mrad. Ferroelectric domain configurations were characterized by vector piezoresponse force microscopy (PFM) using an Asylum Research MFP-3D atomic force microscope equipped with conductive Pt/Ir-coated tips.

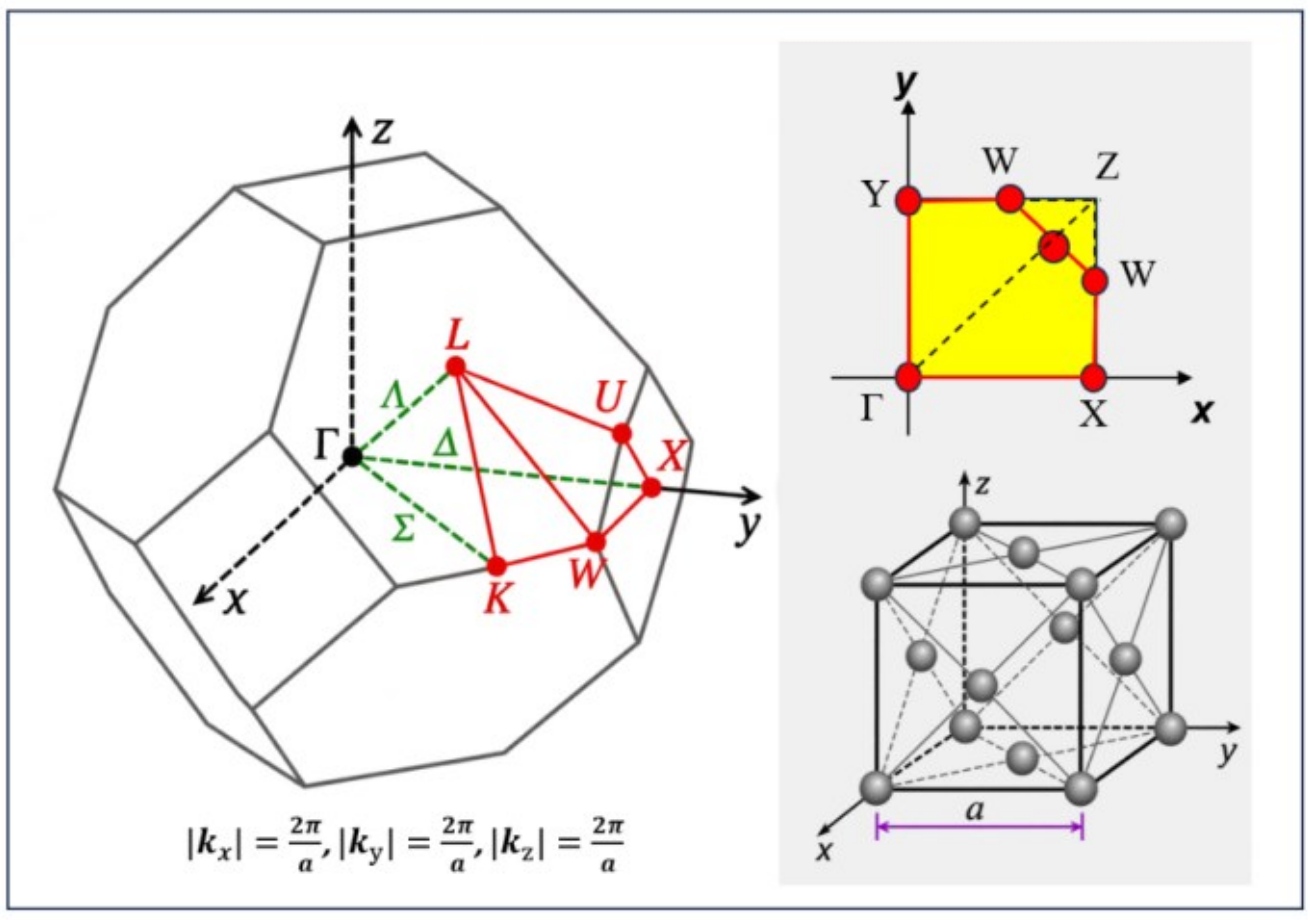

**Fig. S1. Reciprocal-space origin of cubic $HfO_2$.** Brillouin zone of the FCC cubic fluorite reference structure showing the high-symmetry k points and reciprocal-space directions associated with the zone-boundary modes.

The reciprocal-space geometry of cubic $Fm\bar{3}m$ $HfO_2$ provides the foundation for identifying its symmetry-related phonon instabilities. As shown in Fig. S1, the first Brillouin zone of the fluorite structure is that of an fcc lattice (lattice constant $a$) and forms a truncated octahedron bounded by six square and eight hexagonal faces. The square faces satisfy $|k_x| = |k_y| = |k_z| = 2\pi/a$, whereas the hexagonal faces are defined by $|k_x| + |k_y| + |k_z| = 3\pi/a$. In reduced reciprocal coordinates, $k^* = (a/2\pi)k$, the square faces are described by $|k_x^*| = 1$, $|k_y^*| = 1$, and $|k_z^*| = 1$, while the hexagonal faces satisfy $|k_x^*| + |k_y^*| + |k_z^*| = 3/2$. The high-symmetry points of the fcc Brillouin zone, including $\Gamma$, $K$, $L$, $U$, $W$, and $X$, are indicated in Fig. S6. Of particular interest is the $X$ point, located at the square face of the Brillouin zone. In the reduced reciprocal coordinate representation ($|k_x^*| = 1$), its symmetry equivalent realizations can be expressed along the Cartesian basis vectors $(\hat{x}, \hat{y}, \hat{z})$.[6] The corresponding $X$, $Y$, and $Z$ wave vectors X=$\left(0,\frac{1}{2},\frac{1}{2}\right)$, Y=$\left(\frac{1}{2},0,\frac{1}{2}\right)$, Z=$\left(\frac{1}{2},\frac{1}{2},0\right)$ are related by the rotational symmetry of the cubic lattice and therefore represent equivalent directions in reciprocal space. This symmetry equivalence provides the basis for the degenerate phonon modes observed at the corresponding high-symmetry points in the phonon dispersion reveals different unstable branches at these symmetry related wave vectors, including the polar ($\Gamma_{15}^{x}$,$\Gamma_{15}^{y}$,$\Gamma_{15}^{z}$) modes, zone boundary soft modes ($X_2^{\prime x}$, $Y_2^{\prime y}$, and $Z_2^{\prime z}$), and additional ($X_5^{(y,z)}$, $Y_5^{(x,z)}$, $Z_5^{(x,y)}$) modes.

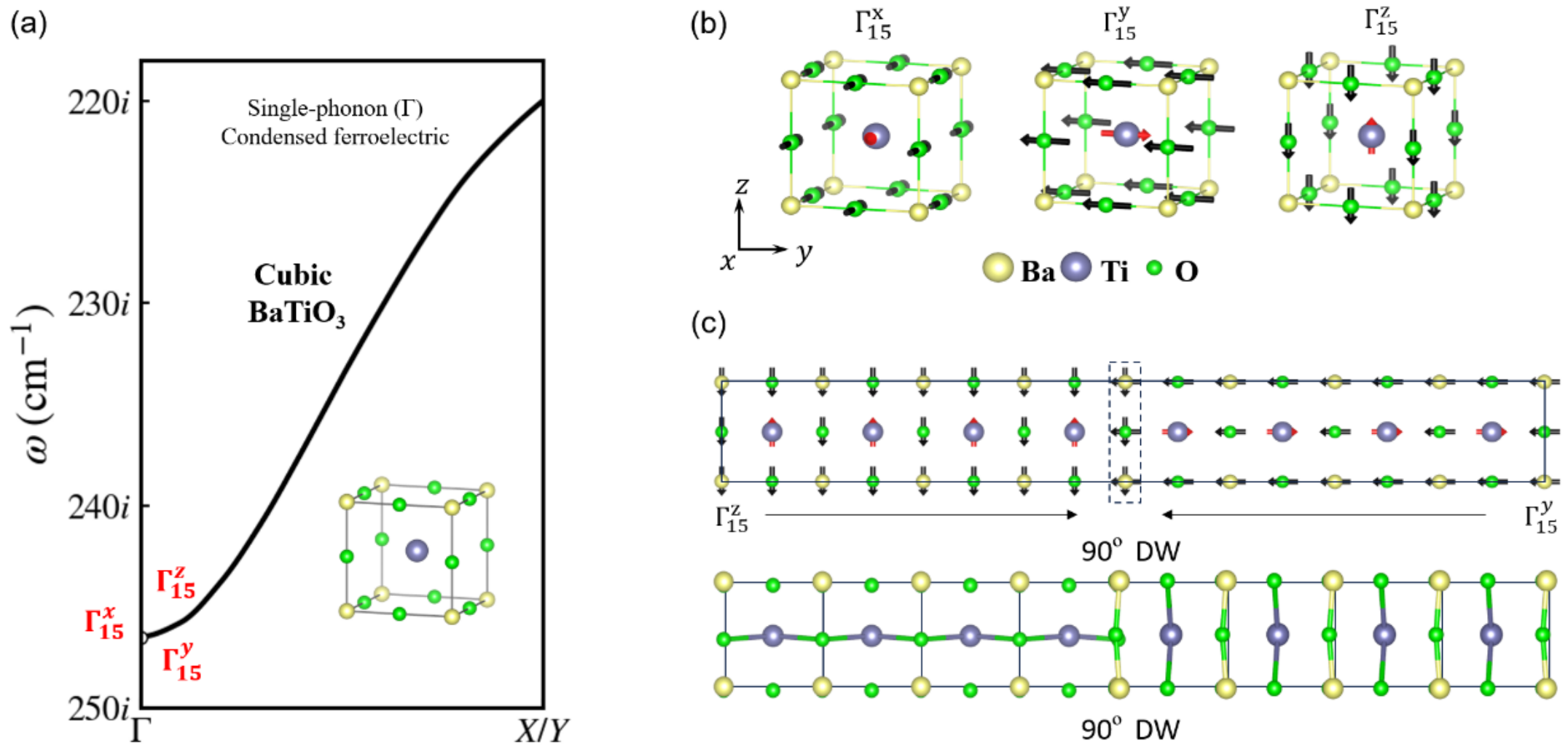

**Fig S2. Mechanism of 90° polarization rotation in BaTiO₃**. (a) Phonon dispersion of instability in cubic $BaTiO_3$ along $\Gamma - X \mid Y$, highlighting the three symmetry-equivalent degenerate components of the unstable $\Gamma_{15}$ polar mode, which condenses to produce the ferroelectric distortion. (b) Atomic displacement patterns of the three $\Gamma_{15}$ polar mode components $\Gamma_{15}^{x}$, $\Gamma_{15}^{y}$, and $\Gamma_{15}^{z}$. (c) Schematic atomic structures of two 90° rotated ferroelectric domains and their intervening domain wall.

### 90° Polarization rotation in BaTiO₃

The microscopic mechanism underlying 90° polarization rotation in $BaTiO_3$ differs fundamentally from that in $HfO_2$ because of the distinct origins of their ferroelectric instabilities. In cubic $BaTiO_3$, ferroelectricity originates from a single unstable zone-center $\Gamma_{15}$ polar mode. Its three degenerate components, $\Gamma_{15}^{x}$, $\Gamma_{15}^{y}$, and $\Gamma_{15}^{z}$ describe symmetry-equivalent polar distortions along the three Cartesian directions. Condensation of any one component breaks the cubic symmetry and establishes a spontaneous polarization along the corresponding direction. Consequently, 90°-rotated ferroelectric variants can be formed by switching the condensed polarization component from one direction to another within the same underlying polar-mode manifold. The rotation therefore does not intrinsically require coupling between distinct phonon modes to establish the ferroelectric phase, although anharmonic mode coupling and elastic interactions can still influence the rotation pathway. Thus, the mechanism at the 90° domain wall requires less energy to form due to single mode change. In $HfO_2$, by contrast, the ferroelectric orthorhombic *Pca2₁* phase cannot be understood simply through the condensation of a single polar mode. Its stabilization involves the cooperative interplay of the Γ-point polar distortion with unstable zone-boundary soft modes and additional dominated lattice distortions. These modes collectively establish the polar structure and its characteristic oxygen sublattice displacements. A 90° polarization rotation therefore involves more than selecting another component of a single polar instability. Hence, it requires the interconversion and reorganization of multiple coupled phonon modes to construct the symmetry equivalent, rotated *$Pca2_1$* variant. The mode amplitudes and displacement patterns must reorganize cooperatively while preserving the essential structural features of the ferroelectric phase.

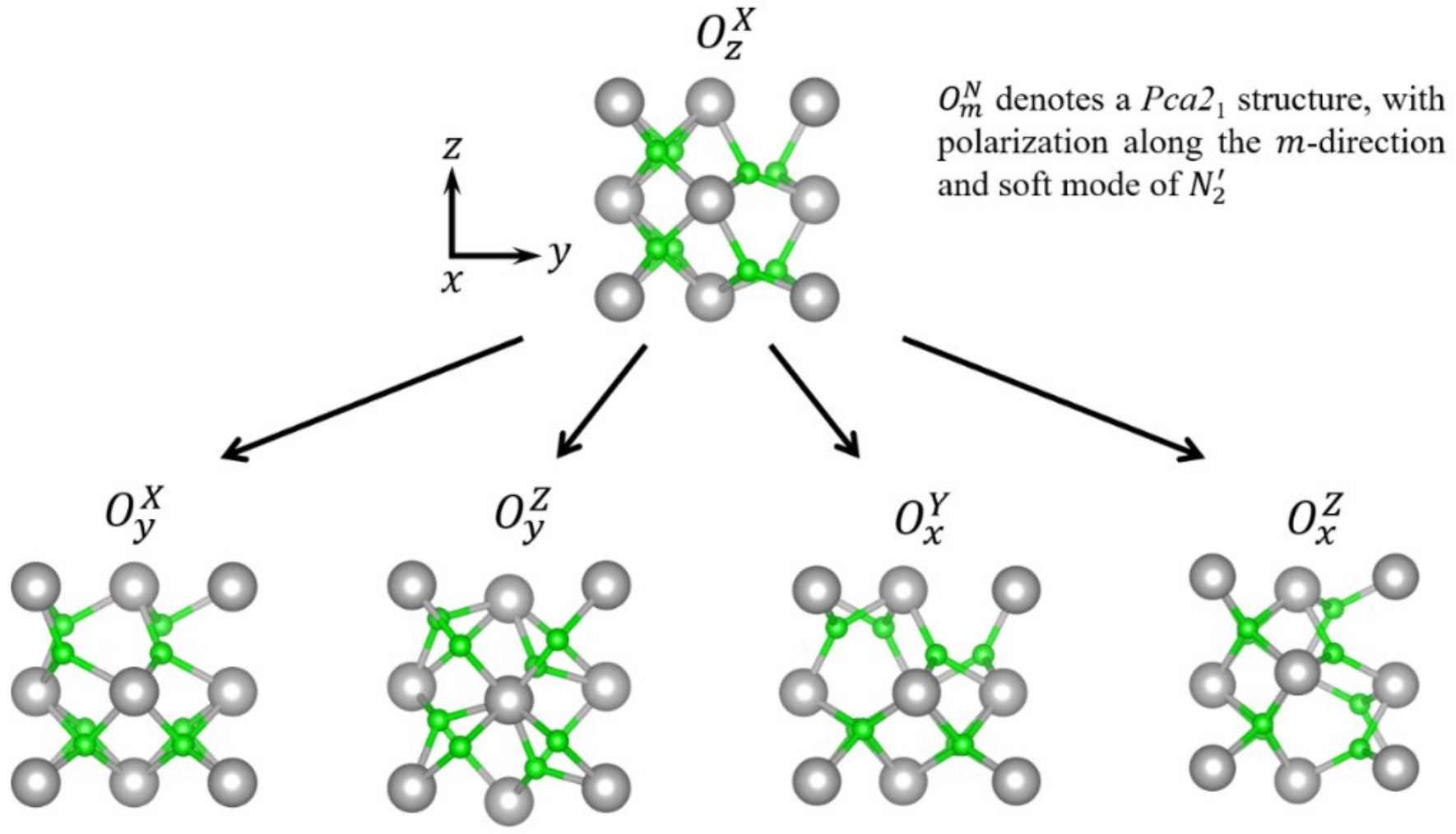

**Fig S3. Schematic of the symmetry-related *$Pca2_1$* phase variants generated from the reference state $\boldsymbol{O_z^X}$**. Here, denotes a structure with polarization along the direction and the soft mode oriented along. The four resulting configurations, $O_y^X$, $O_y^Z$, $O_x^Y$, and $O_x^Z$ illustrate the distinct combinations of polar and anti-polar-mode orientations associated with polarization rotation.

**Table S1. 5 variants of *O* phases of $HfO_2$.** Computed amplitudes (Å) of each phonon mode of structures are listed in the table. Oxygen modes and hafnia modes are separate for clarity.

| Structure | | $\Gamma_{15}^x$ | $\Gamma_{15}^y$ | $\Gamma_{15}^z$ | $X_2^{\prime x}$ | $Y_2^{\prime y}$ | $Z_2^{\prime z}$ | $X_5^y$ | $X_5^z$ | $Y_5^x$ | $Y_5^z$ | $Z_5^x$ | $Z_5^y$ | $X_5^{\prime y}$ | $X_5^{\prime z}$ | $Y_5^{\prime x}$ | $Y_5^{\prime z}$ | $Z_5^{\prime x}$ | $Z_5^{\prime y}$ |
|---|---|---|---|---|---|---|---|---|---|---|---|---|---|---|---|---|---|---|---|
| $\boldsymbol{O_z^X}$ | Q(Å) | 0 | 0 | 0.402 | 0.488 | 0 | 0 | 0.397 | 0 | 0 | 0.394 | 0.377 | 0 | 0 | 0 | 0 | 0 | 0 | 0.110 |
| $\boldsymbol{O_y^X}$ | Q(Å) | 0 | 0.402 | 0 | 0.488 | 0 | 0 | 0 | 0.397 | 0.377 | 0 | 0 | 0.394 | 0 | 0 | 0 | 0.110 | 0 | 0 |
| $\boldsymbol{O_y^Z}$ | Q(Å) | 0 | 0.402 | 0 | 0 | 0 | 0.488 | 0.394 | 0 | 0 | 0.377 | 0.397 | 0 | 0 | 0 | 0.110 | 0 | 0 | 0 |
| $\boldsymbol{O_x^Y}$ | Q(Å) | 0.402 | 0 | 0 | 0 | 0.488 | 0 | 0.377 | 0 | 0 | 0.397 | 0.394 | 0 | 0 | 0.110 | 0 | 0 | 0 | 0 |
| $\boldsymbol{O_x^Z}$ | Q(Å) | 0.402 | 0 | 0 | 0 | 0 | 0.488 | 0 | 0.377 | 0.394 | 0 | 0 | 0.397 | 0.110 | 0 | 0 | 0 | 0 | 0 |

| Structure | | $X_3^{\prime x}$ | $Y_3^{\prime y}$ | $Z_3^{\prime z}$ | $X_5^{\prime y}$ | $X_5^{\prime z}$ | $Y_5^{\prime x}$ | $Y_5^{\prime z}$ | $Z_5^{\prime x}$ | $Z_5^{\prime y}$ |
|---|---|---|---|---|---|---|---|---|---|---|
| $\boldsymbol{O_z^X}$ | Q(Å) | 0 | 0.089 | 0 | 0 | 0 | 0 | 0 | 0.173 | 0 |
| $\boldsymbol{O_y^X}$ | Q(Å) | 0 | 0 | 0.089 | 0 | 0 | 0.173 | 0 | 0 | 0 |
| $\boldsymbol{O_y^Z}$ | Q(Å) | 0.089 | 0 | 0 | 0 | 0 | 0 | 0.173 | 0 | 0 |
| $\boldsymbol{O_x^Y}$ | Q(Å) | 0 | 0 | 0.089 | 0.173 | 0 | 0 | 0 | 0 | 0 |
| $\boldsymbol{O_x^Z}$ | Q(Å) | 0 | 0.089 | 0 | 0 | 0.173 | 0 | 0 | 0 | 0 |

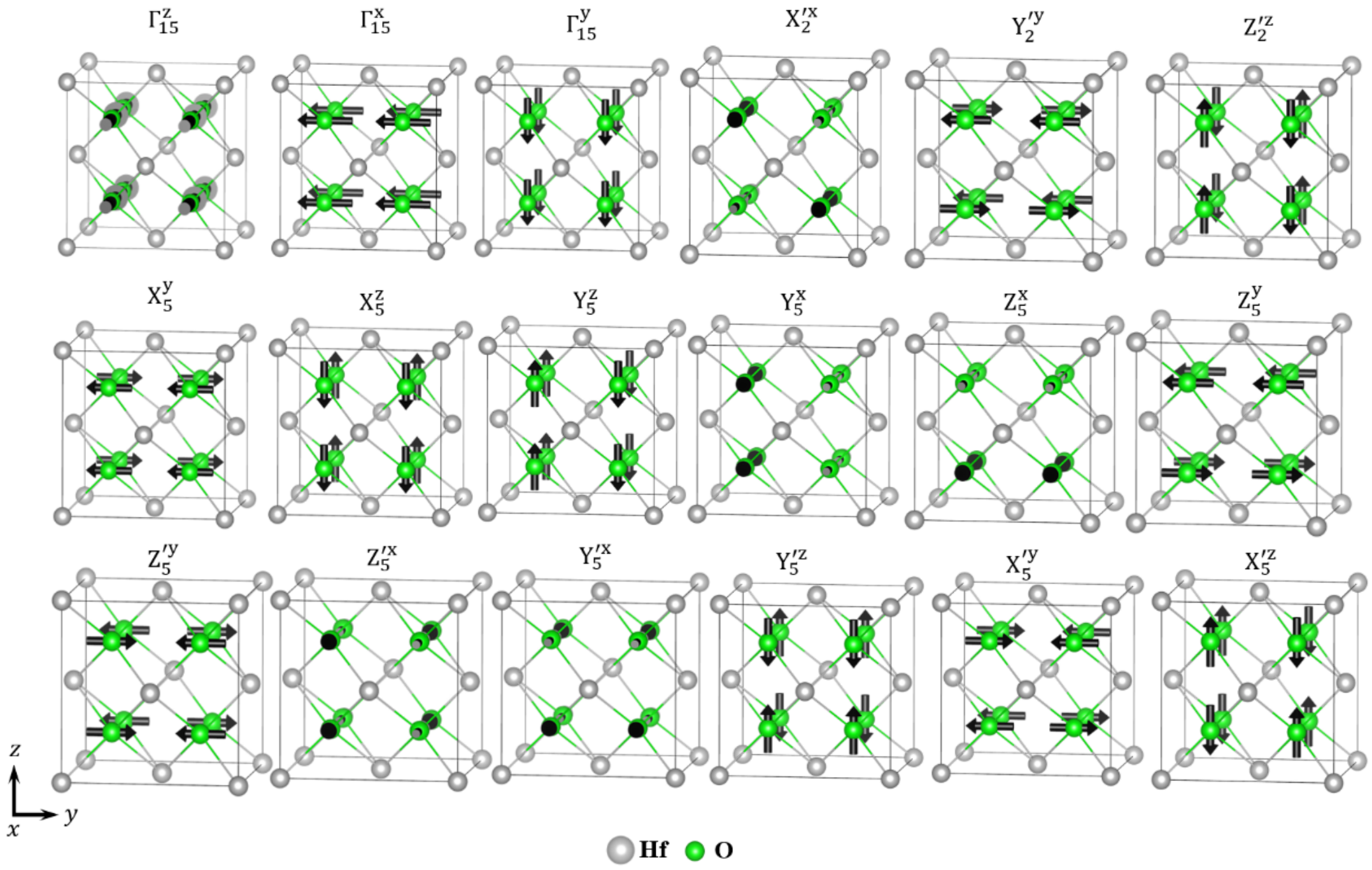

**Fig S4. Structural visualization of oxygen phonon modes associated with the orthorhombic *Pca2₁* phase of $HfO_2$ and its rotated variants with the cubic phase as reference**. Grey and green spheres represent Hf and O atoms, respectively, while arrows indicate the direction and relative amplitude of the corresponding atomic displacements. The visualized modes capture the distinct polar and nonpolar distortions that characterize the o-phase and provide the microscopic structural basis for its polarization and mode-coupling behavior.

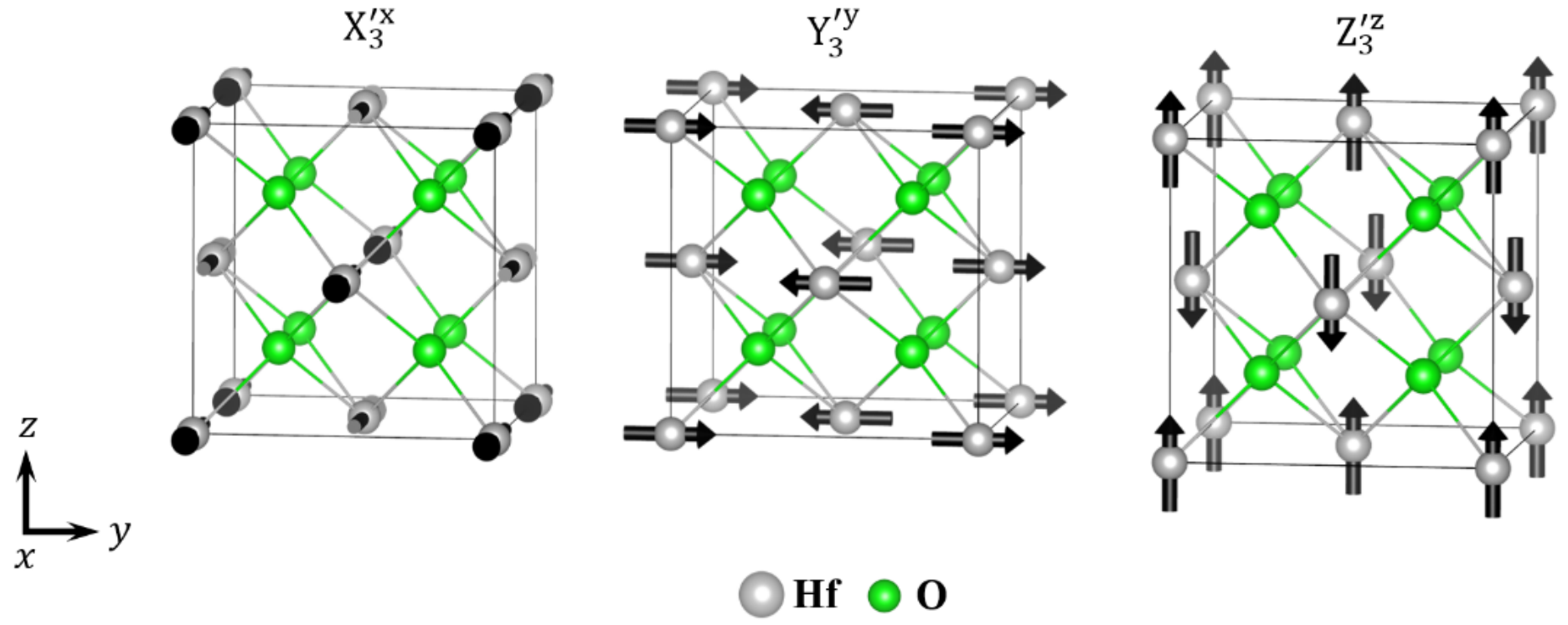

**Fig S5. Structural visualization of hafnia phonon modes associated with the orthorhombic $Pca2_1$ phase of $HfO_2$ and its rotated variants with the cubic phase as reference**. Grey and green spheres represent Hf and O atoms, respectively, while arrows indicate the direction and relative amplitude of the corresponding atomic displacements. The visualized modes capture the distinct polar and nonpolar distortions that characterize the *$Pca2_1$* phase and provide the microscopic structural basis for its polarization and mode-coupling behavior.

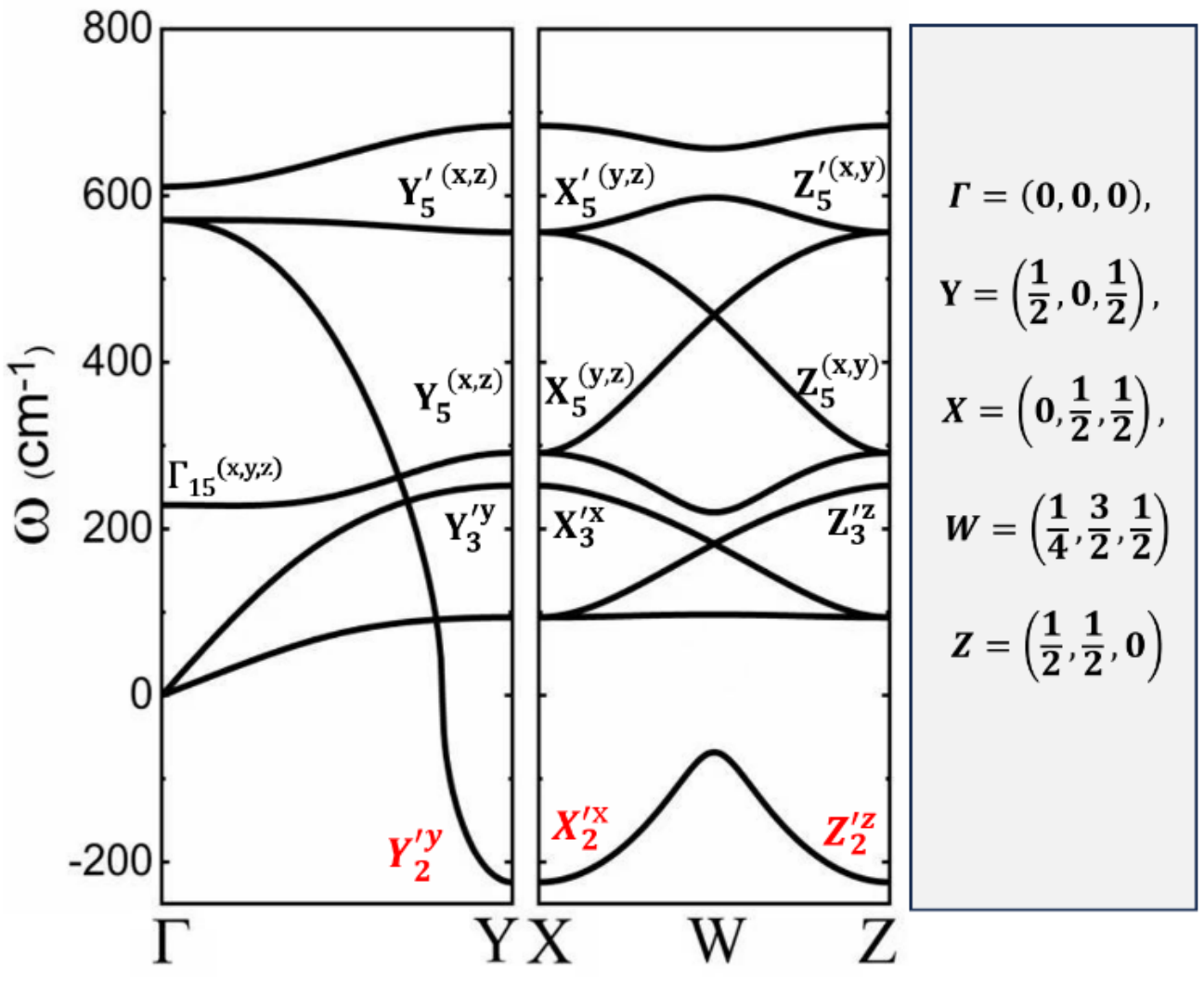

**Fig S6. Phonon dispersion of the orthorhombic phase of $HfO_2$ along the high-symmetry path Γ−Y | X−W−Z**. The branches identify the symmetry-labeled phonon modes relevant to the structural distortions of the *$Pca2_1$* phase, including the zone-center polar mode and the zone-boundary modes. The imaginary frequency indicated in red represent the zone boundary soft modes instability associated with the phase transition distortion from nonpolar and polar phase.

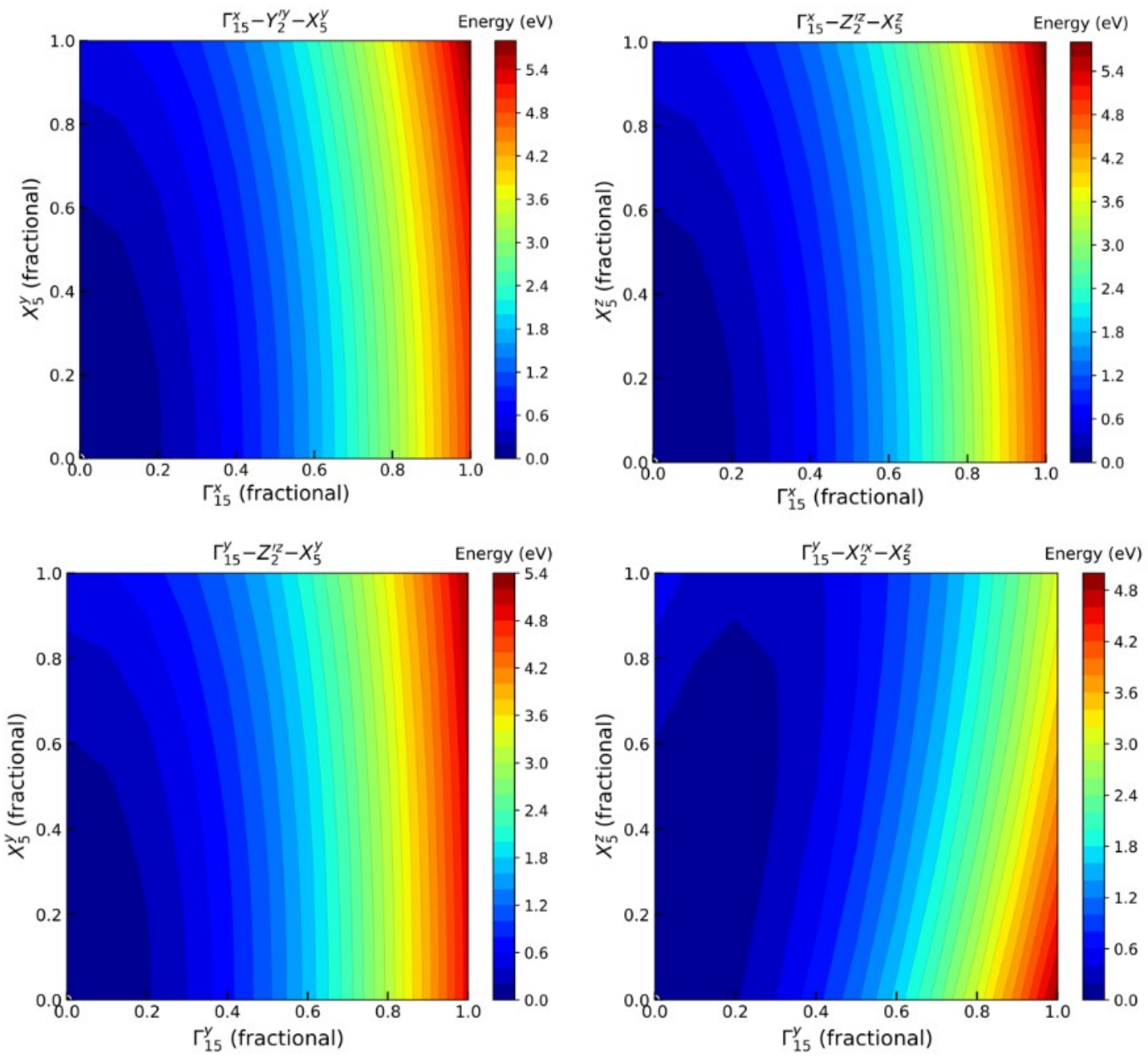

**Fig S7. Two-dimensional energy map of trilinear coupled lattice-distortion modes**. (a) Energy contour map of $\Gamma_{15}^{x}$ and $X_5^y$ with the amplitude of $Y_2^{\prime y}$ fixed at its maximum value. (b) Energy of $\Gamma_{15}^{x}$ and $X_5^z$ with the amplitude of $Z_2^{\prime z}$ fixed at its maximum value. (c) Energy of $\Gamma_{15}^{y}$ and $X_5^y$ with the amplitude of $Z_2^{\prime z}$ fixed at its maximum value. (d) Energy of $\Gamma_{15}^{y}$ and $X_5^z$ with the amplitude of $X_2^{\prime x}$ fixed at its maximum value.

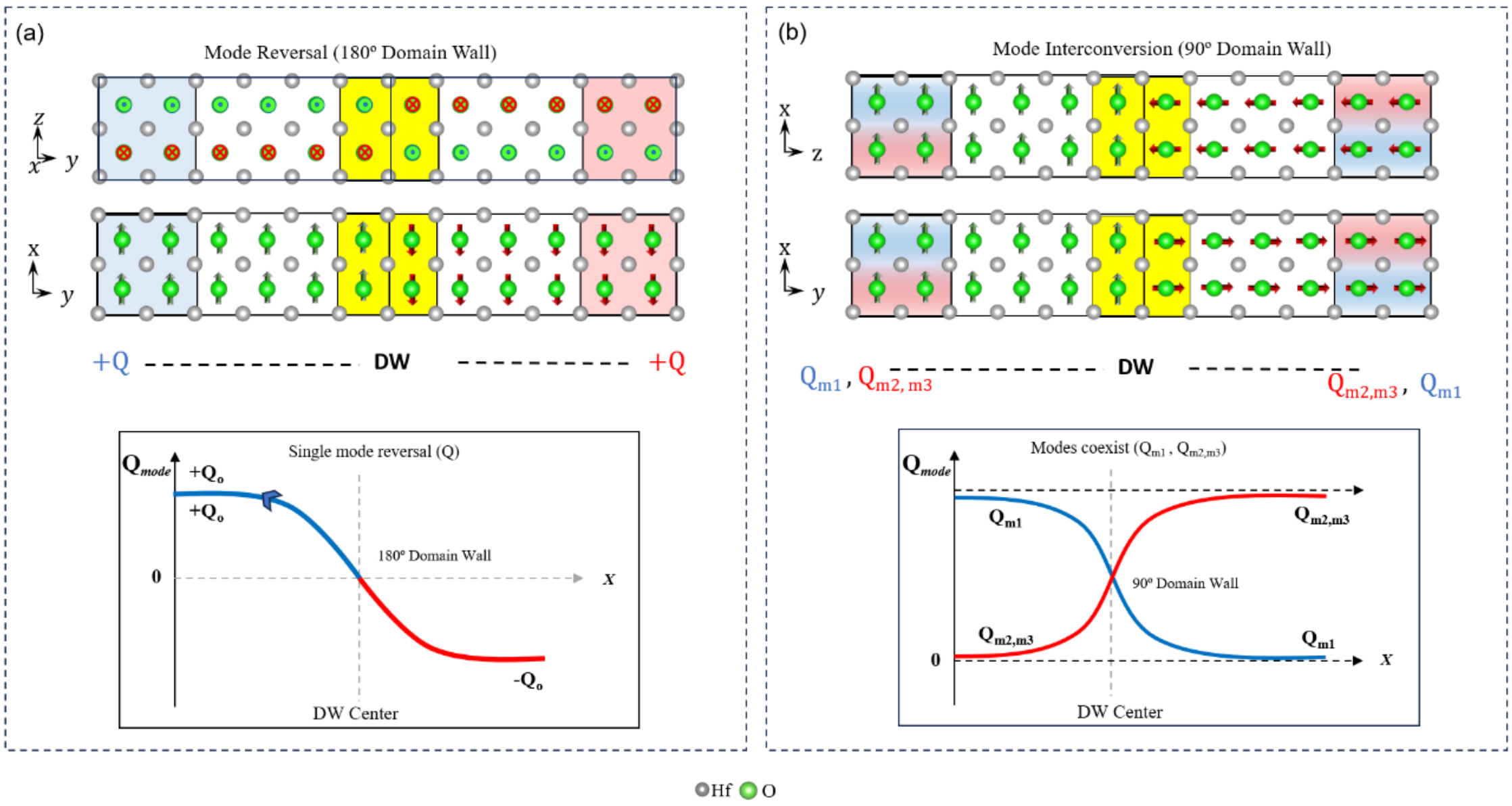

**FIG S8.** Domain wall Mode evolution. (a) 180° domain wall mode evolution, where mode across the domain undergoes mode sign change, with mode at the domain center exerting no amplitude. (b) 90° domain wall mode evolution, where the modes across the domain interconvert and coexist at the domain wall center with substantial amplitude.

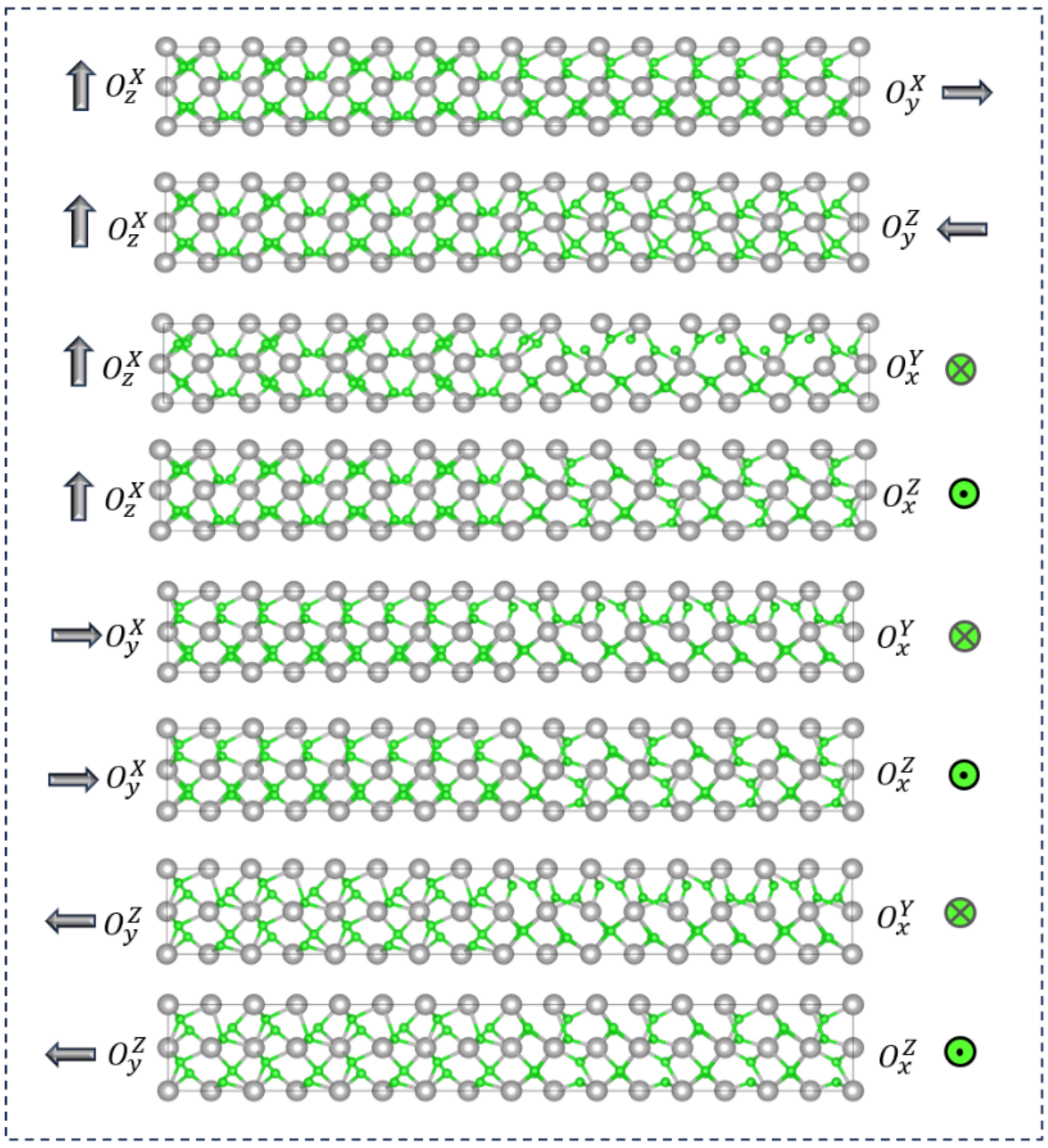

**Fig S9. 90° Domain wall variants of *Pca2₁* phase of HfO₂**. The distinct combinations of variant 90° single domain to form domain walls. The labels and symbols denote the respective polarization and mode directions. The configurations illustrate how polarization rotation is accompanied by a corresponding reorientation of the structural mode.

**Table S2.** Calculated domain-wall energies for different 90° domain configurations in orthorhombic $Pca2_1$ of $HfO_2$ using ***Hf_sv Standard potential***. The values represent the energy cost of forming domain walls between different polarization variants.

| **Domain wall** | **Polarization** | **Charge state** | **DWE ($mJm^{-2}$)** |
|---|---|---|---|
| $O_z^X/O_y^X$ | $+z \rightarrow +y$ | Charged | 567.1 |
| $O_z^X/O_y^Z$ | $+z \rightarrow -y$ | Charged | 674.3 |
| $O_z^X/O_x^Y$ | $+z \rightarrow -x$ | Neutral | 484.7 |
| $O_z^X/O_x^Z$ | $+z \rightarrow +x$ | Neutral | 469.6 |
| $O_y^X/O_x^Y$ | $+y \rightarrow -x$ | Charged | 779.8 |
| $O_y^X/O_x^Z$ | $+y \rightarrow +x$ | Charged | 663.7 |
| $O_y^Z/O_x^Y$ | $-y \rightarrow -x$ | Charged | 834.6 |
| $O_y^Z/O_x^Z$ | $-y \rightarrow +x$ | Charged | 570.0 |

The Domain Wall Energy (DWE) is analyzed to determines the stability of the 90° domain walls which expressed as

$$DWE = \frac{E_{DW} - E_{bulk}}{2A}$$

with and denoting the total energies of the domain wall $E_{DW}$ and corresponding bulk supercells $E_{bulk}$, respectively, and the cross-sectional area of a single domain wall $A$.

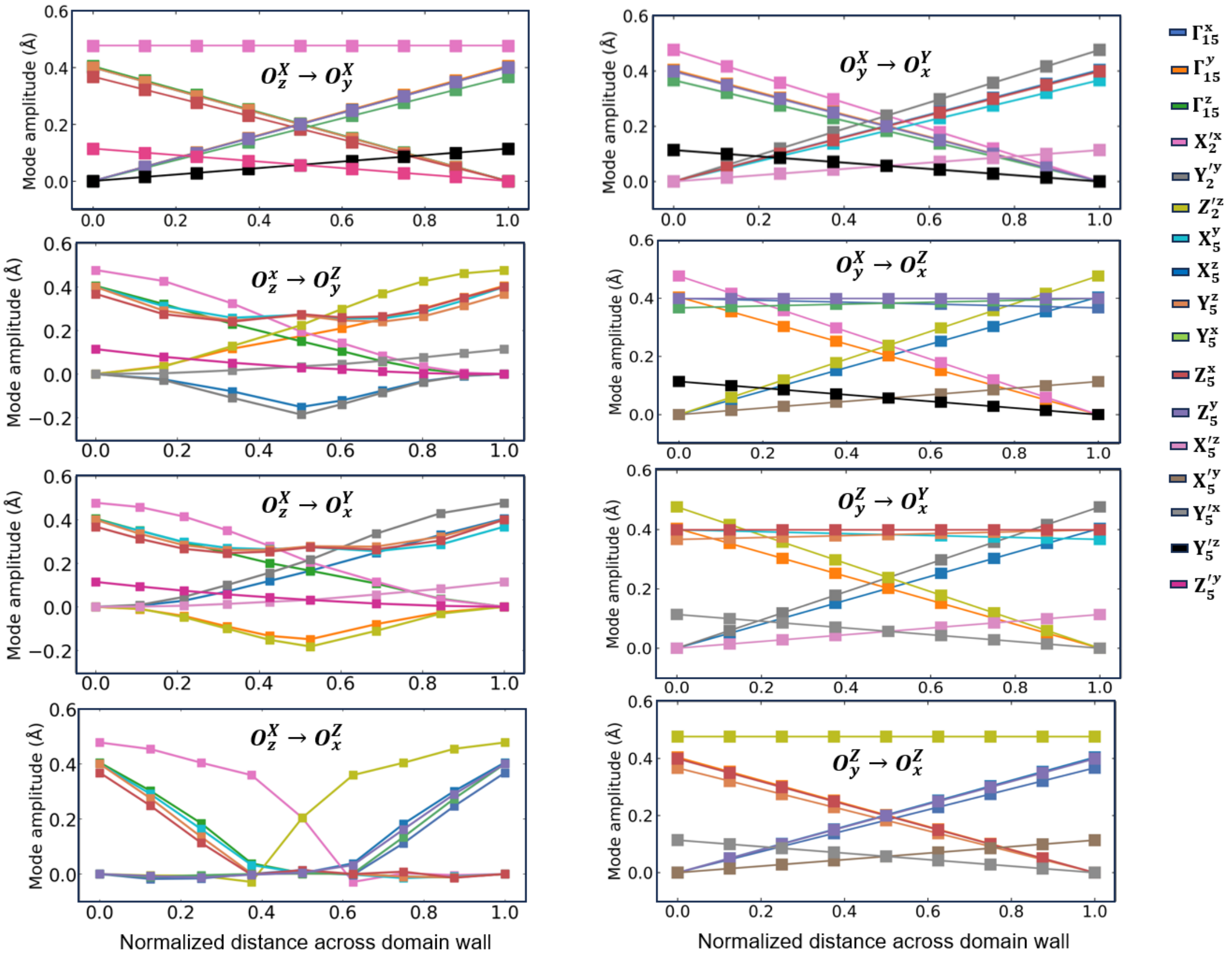

**Fig S10.** Spatial evolution of the symmetry mode amplitudes across all generated 90º domain walls. The units of these structural order parameters are that of distance (in Å) because they are measuring the amount of distortion.

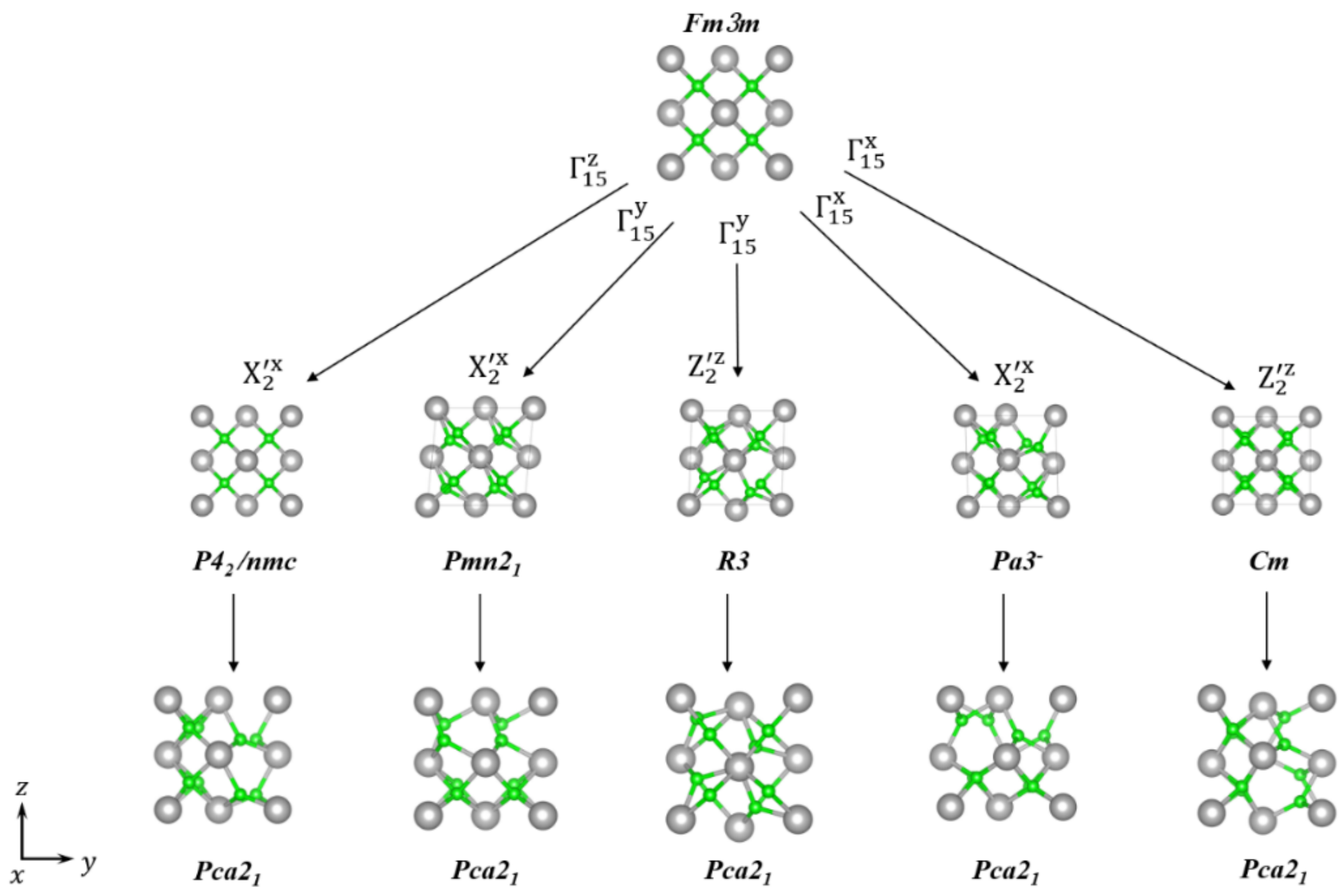

**Fig S11.** Structural transformation from the parent cubic phase to orthorhombic *Pca2₁* variants of $HfO_2$ with their respective intermediate phases and dominant modes.

The coexistence of multiple lattice-distortion modes is dictated by symmetry-allowed interactions in the free energy, with trilinear coupling playing a particularly important role in $HfO_2$. To investigate the energetic interactions among the relevant phonon modes associated with the rotated orthorhombic variants of fluorite $HfO_2$, we construct a Landau-type energy expansion in terms of the corresponding symmetry-mode amplitudes. The cubic fluorite phase $(Fm\bar{3}m)$ serves as the reference structure. Five rotated variants, denoted $(O_z^X), (O_y^X), (O_y^Z), (O_x^Y)$, and $(O_x^Z)$, are considered. For each variant, the energy is expressed as a function of eight mode amplitudes, including the zone-center polar mode and selected zone-boundary modes. $\{\Gamma_{15}^{z}, X_2^{\prime x}, X_5^{y}, Y_5^{z}, Z_5^{x}, Z_5^{\prime y}, Z_5^{\prime x}, Y_3^{\prime y}\}$, $\{\Gamma_{15}^{y}, X_2^{\prime x}, X_5^{z}, Y_5^{x}, Z_5^{y}, Y_5^{\prime x}, Y_5^{\prime z}, Z_3^{\prime z}\}$, $\{\Gamma_{15}^{y}, Z_2^{\prime z}, X_5^{y}, Y_5^{z}, Z_5^{x}, Y_5^{\prime x}, Y_5^{\prime z}, X_3^{\prime x}\}$, $\{\Gamma_{15}^{x}, Y_2^{\prime y}, X_5^{y}, Y_5^{z}, Z_5^{x}, X_5^{\prime y}, X_5^{\prime z}, Z_3^{\prime z}\}$, and $\{\Gamma_{15}^{x}, Z_2^{\prime z}, X_5^{z}, Y_5^{x}, Z_5^{y}, X_5^{\prime y}, X_5^{\prime z}, Y_3^{\prime y}\}$ corresponding to $O_z^X$, $O_y^X$, $O_y^Z$, $O_x^Y$, and $O_x^Z$ respectively.

The allowed trilinear coupling terms are expressed as the following Equation.

(a) $O_z^X$

$$
\begin{aligned}
E_{\mathrm{tri}}^{(O_z^X)} = {} & a_1 Q_{\Gamma_{15}^z} Q_{X_2^{\prime x}} Q_{X_5^y} + a_2 Q_{\Gamma_{15}^z} Q_{Y_3^{\prime y}} Q_{Y_5^z} \\
& + a_3 Q_{\Gamma_{15}^z} Q_{Z_5^{\prime y}} Q_{Z_5^x} + a_4 Q_{\Gamma_{15}^z} Q_{Z_5^{\prime x}} Q_{Z_5^x} \\
& + a_5 Q_{X_2^{\prime x}} Q_{Y_3^{\prime y}} Q_{Z_5^x} + a_6 Q_{X_2^{\prime x}} Q_{Y_5^z} Q_{Z_5^{\prime y}} \\
& + a_7 Q_{X_2^{\prime x}} Q_{Y_5^z} Q_{Z_5^{\prime x}} + a_8 Q_{X_5^y} Q_{Y_5^z} Q_{Z_5^x} \\
& + a_9 Q_{X_5^y} Q_{Y_3^{\prime y}} Q_{Z_5^{\prime y}} + a_{10} Q_{X_5^y} Q_{Y_3^{\prime y}} Q_{Z_5^{\prime x}}
\end{aligned}
$$

(b) $O_y^X$

$$
\begin{aligned}
E_{\mathrm{tri}}^{(O_y^X)} = {} & a_1 Q_{\Gamma_{15}^y} Q_{X_2^{\prime x}} Q_{X_5^z} + a_2 Q_{\Gamma_{15}^y} Q_{Z_3^{\prime z}} Q_{Y_5^x} \\
& + a_3 Q_{\Gamma_{15}^y} Q_{Y_5^{\prime x}} Q_{Z_5^y} + a_4 Q_{\Gamma_{15}^y} Q_{Y_5^{\prime z}} Q_{Z_5^y} \\
& + a_5 Q_{X_2^{\prime x}} Q_{Z_3^{\prime z}} Q_{Z_5^y} + a_6 Q_{X_2^{\prime x}} Q_{Y_5^x} Q_{Y_5^{\prime x}} \\
& + a_7 Q_{X_2^{\prime x}} Q_{Y_5^x} Q_{Y_5^{\prime z}} + a_8 Q_{X_5^z} Q_{Y_5^x} Q_{Z_5^y} \\
& + a_9 Q_{X_5^z} Q_{Z_3^{\prime z}} Q_{Y_5^{\prime x}} + a_{10} Q_{X_5^z} Q_{Z_3^{\prime z}} Q_{Y_5^{\prime z}}
\end{aligned}
$$

(c) $O_y^Z$

$$
\begin{aligned}
E_{\mathrm{tri}}^{(O_y^Z)} = {} & a_1 Q_{\Gamma_{15}^y} Q_{Z_2^{\prime z}} Q_{X_5^y} + a_2 Q_{\Gamma_{15}^y} Q_{X_3^{\prime x}} Q_{Y_5^z} \\
& + a_3 Q_{\Gamma_{15}^y} Q_{Y_5^{\prime x}} Q_{Z_5^x} + a_4 Q_{\Gamma_{15}^y} Q_{Y_5^{\prime z}} Q_{Z_5^x} \\
& + a_5 Q_{Z_2^{\prime z}} Q_{X_3^{\prime x}} Q_{Z_5^x} + a_6 Q_{Z_2^{\prime z}} Q_{Y_5^z} Q_{Y_5^{\prime x}} \\
& + a_7 Q_{Z_2^{\prime z}} Q_{Y_5^z} Q_{Y_5^{\prime z}} + a_8 Q_{X_5^y} Q_{Y_5^z} Q_{Z_5^x} \\
& + a_9 Q_{X_5^y} Q_{X_3^{\prime x}} Q_{Y_5^{\prime x}} + a_{10} Q_{X_5^y} Q_{X_3^{\prime x}} Q_{Y_5^{\prime z}}
\end{aligned}
$$

(d) $O_x^Y$

$$
\begin{aligned}
E_{\mathrm{tri}}^{(O_x^Y)} = {} & a_1 Q_{\Gamma_{15}^x} Q_{Y_2^{\prime y}} Q_{X_5^y} + a_2 Q_{\Gamma_{15}^x} Q_{Z_3^{\prime z}} Q_{Y_5^z} \\
& + a_3 Q_{\Gamma_{15}^x} Q_{X_5^{\prime y}} Q_{Z_5^x} + a_4 Q_{\Gamma_{15}^x} Q_{X_5^{\prime z}} Q_{Z_5^x} \\
& + a_5 Q_{Y_2^{\prime y}} Q_{Z_3^{\prime z}} Q_{Z_5^x} + a_6 Q_{Y_2^{\prime y}} Q_{Y_5^z} Q_{X_5^{\prime y}} \\
& + a_7 Q_{Y_2^{\prime y}} Q_{Y_5^z} Q_{X_5^{\prime z}} + a_8 Q_{X_5^y} Q_{Y_5^z} Q_{Z_5^x} \\
& + a_9 Q_{X_5^y} Q_{Z_3^{\prime z}} Q_{X_5^{\prime y}} + a_{10} Q_{X_5^y} Q_{Z_3^{\prime z}} Q_{X_5^{\prime z}}
\end{aligned}
$$

(e) $O_x^Z$

$$
\begin{aligned}
E_{\mathrm{tri}}^{(O_x^Z)} = {} & a_1 Q_{\Gamma_{15}^x} Q_{Z_2{}'^z} Q_{X_5^z} + a_2 Q_{\Gamma_{15}^x} Q_{Y_3{}'^y} Q_{Y_5^x} \\
& + a_3 Q_{\Gamma_{15}^x} Q_{X_5{}'^y} Q_{Z_5^y} + a_4 Q_{\Gamma_{15}^x} Q_{X_5{}'^z} Q_{Z_5^y} \\
& + a_5 Q_{Z_2{}'^z} Q_{Y_3{}'^y} Q_{Z_5^y} + a_6 Q_{Z_2{}'^z} Q_{Y_5^x} Q_{X_5{}'^y} \\
& + a_7 Q_{Z_2{}'^z} Q_{Y_5^x} Q_{X_5{}'^z} + a_8 Q_{X_5^z} Q_{Y_5^x} Q_{Z_5^y} \\
& + a_9 Q_{X_5^z} Q_{Y_3{}'^y} Q_{X_5{}'^y} + a_{10} Q_{X_5^z} Q_{Y_3{}'^y} Q_{X_5{}'^z}.
\end{aligned}
$$

Here, $\left(E_{\mathrm{tri}}^{(v)}\right)$ denotes the trilinear coupling contribution to the Landau free energy for the polarization variant $(v)$, where $\left(v \in O_z^X, O_y^X, O_y^Z, O_x^Y, O_x^Z\right)$. The quantities $(Q_\lambda^\alpha)$ represent the amplitudes of symmetry-adapted phonon modes, with $(\lambda)$ identifying the corresponding phonon-mode irreducible representation or Brillouin-zone wavevector label (e.g., $(\Gamma_{15}), (X_5), (Y_5)$, and $(Z_5)$), and $(\alpha)$ denoting the Cartesian component of the mode. The coefficients $(a_i)((i = 1, \ldots, 10))$ quantify the strengths of the respective trilinear interactions among three phonon-mode amplitudes. Each product describes a coupling between three modes that contribute to the stabilization of a particular structural variant.

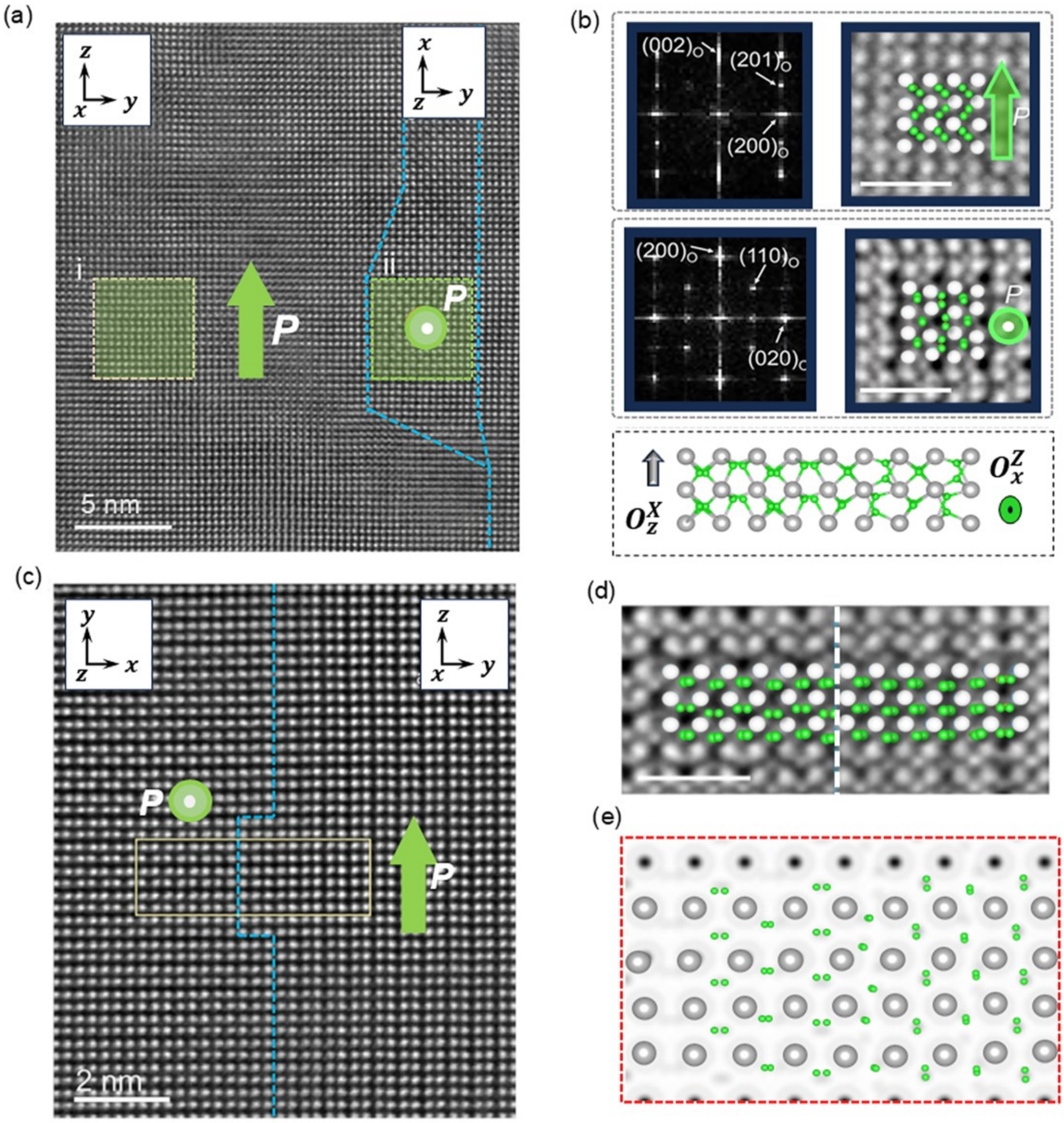

**FIG 12**. Experimental observation of 90° domain wall in $HfO_2$. (a) Atomic-resolution scanning transmission electron microscopy (STEM) image showing 5 nm resolution of two ferroelectric domains with regions forming 90° orientation matching to $O_z^X/O_x^Z$. (b) Corresponding Fourier-transform patterns and atomic-resolution images from the two domain regions, illustrating their distinct crystallographic orientations. Structural model illustrating $O_z^X$ / $O_x^Z$ domain. (c) Atomic-resolution scanning transmission electron microscopy (STEM) image showing 2 nm resolution of 90° $O_z^X/O_x^Z$ ferroelectric domains (d) Enlarged atomic-resolution images across the domain wall, highlighting the local atomic arrangement on either side of the interface. (e) Simulated Annular Bright Field (ABF) images of the $O_z^X/O_x^Z$ domain structure.